\documentclass{aa}  

\usepackage{graphicx}
\usepackage{txfonts}
\usepackage{graphics,graphicx}
\usepackage{amsmath}
\usepackage{amssymb}
\usepackage{tabularx}
\usepackage{graphics,graphicx}
\usepackage{txfonts}
\usepackage{longtable}
\usepackage{textcomp}
\usepackage{hhline}
\usepackage{arydshln}
\usepackage{multirow}
\usepackage{lscape}
\usepackage{array}
\usepackage{natbib}
\usepackage{setspace}
\usepackage{multicol}
\usepackage{pifont} 
\usepackage{wrapfig}
\usepackage{amsmath}
\usepackage{hyperref}
\usepackage{fancybox}
\usepackage{color}
\usepackage{subcaption}
\usepackage[dvipsnames]{xcolor}
\usepackage[export]{adjustbox}
\usepackage[normalem]{ulem}
\usepackage[separate-uncertainty=true,multi-part-units=single]{siunitx}
\usepackage{acronym}
\usepackage[capitalise]{cleveref}
\usepackage{placeins}
\usepackage{float}

\usepackage{titlesec}

\makeatletter
\newlength{\sfp@hseplen}\newlength{\sfp@vseplen}
\define@cmdkey{subfigpos}[sfp@]{pos}[ul]{}
\define@cmdkey{subfigpos}[sfp@]{font}[\small]{}
\define@cmdkey{subfigpos}[sfp@]{vsep}[2\baselineskip]{\setlength{\sfp@vseplen}{\sfp@vsep}}
\define@cmdkey{subfigpos}[sfp@]{hsep}[10pt]{\setlength{\sfp@hseplen}{\sfp@hsep}}
\newcommand{\subfigimg}[4][,]{%
  \setkeys{subfigpos}{pos,font,vsep,hsep,#2}
  \setbox1=\hbox{\adjincludegraphics[#1]{#4}}
  \ifnum\pdfstrcmp{\sfp@pos}{ul}=0
    \leavevmode\usebox1%
    \rlap{\hspace*{\sfp@hsep}\raisebox{\dimexpr\ht1-\sfp@vsep}[0pt][0pt]{\sfp@font{#3}}}
  \else\ifnum\pdfstrcmp{\sfp@pos}{ur}=0
    \leavevmode\usebox1%
    \llap{\raisebox{\dimexpr\ht1-\sfp@vsep}[0pt][0pt]{\sfp@font{#3}}\hspace*{\sfp@hsep}}
  \else\ifnum\pdfstrcmp{\sfp@pos}{lr}=0
    \leavevmode\usebox1%
    \llap{\raisebox{\sfp@vsep}[0pt][0pt]{\sfp@font{#3}}\hspace*{\sfp@hsep}}
  \else
    \leavevmode\rlap{\usebox1}
    \rlap{\hspace*{\sfp@hseplen}\raisebox{\sfp@vsep}[0pt][0pt]{\sfp@font{#3}}}
    \phantom{\usebox1}
  \fi\fi\fi
}
\makeatother

\DeclareSIUnit[]\hcpu{\text{h\ensuremath{_{\rm cpu}}}}

\newacro{nir}[NIR]{near-infrared}
\newacro{mir}[MIR]{mid-infrared}
\newacro{fir}[FIR]{far-infrared}
\newacro{mm}[mm-range]{millimetre wavelength range}
\newacro{cm}[CM]{core-mantle}
\newacro{emt}[EMT]{effective medium theory}
\newacro{dda}[DDA]{discrete dipole approximation}
\newacro{mhd}[MHD]{magnetohydrodynamics}
\newacro{ism}[ISM]{interstellar medium}
\newacro{amr}[AMR]{adaptative mesh refinement}
\newacro{los}[LOS]{line of sight}
\newacro{rt}[RT]{radiative transfer}
\newacro{mft}[MFT]{mean field theory}
\newacro{tmm}[TMM]{T-Matrix method}
\newacro{mmf}[MMF]{modified mean field theory}
\newacro{dhs}[DHS]{distribution of hollow spheres}
\newacro{omer}[OMER]{optimisation method by equivalent representation}

\def\eff{\mathrm{eff}}
\def\abs{\mathrm{abs}}
\def\back{\mathrm{back}}
\def\sca{\mathrm{sca}}
\def\ext{\mathrm{ext}}

\def\opt{\mathrm{opt}}
\def\obj{\mathrm{obj}}
\def\d{\mathrm{d}}

\begin{document}

   \title{Computation of complex dust grain optical properties: Optimisation method by equivalent representation (OMER)}
   \titlerunning{Complex dust grain optical properties: Optimisation method by equivalent representation (OMER)}

   \subtitle{}

   \author{M.-A. Carpine\inst{1} \and
           N. Ysard\inst{2,3} \and A. Maury\inst{4,5,1}}

   \institute{ Université Paris-Saclay, Université Paris Cité, CEA, CNRS, AIM, 91191, Gif-sur-Yvette, France
         \\
         \email{marie-anne.carpine@cea.fr}
         \and IRAP, CNRS, Université de Toulouse, 9 avenue du Colonel Roche, 31028 Toulouse Cedex 4, France
         \and Universit{\'e} Paris-Saclay, CNRS, Institut d'Astrophysique Spatiale, 91405, Orsay, France
         \and
         Institute of Space Sciences (ICE), CSIC, Campus UAB, Carrer de Can Magrans s/n, E-08193 Barcelona, Spain
         \and ICREA, Pg. Lluís Companys 23, Barcelona, Spain}

   \date{\today}

  \abstract
   {Computing the optical properties of dust grains is important for understanding various astrophysical environments. However, the numerical methods used to compute these properties require significant computing resources for irregularly shaped or inhomogeneous particles that are representative of realistic grown dust.}
   {This study aims to provide an efficient yet accurate \ac{dda} computation method for large aggregates with 'complex' monomers and structural details or compositional inhomogeneities.}
   {We propose the \ac{omer} method, which uses optimisation of the optical properties of a known monomer to find an equivalent homogeneous sphere. This sphere can then be used as a substitute for the complex monomer in large aggregates.}
   {The method provides a good approximation of complex monomers, as demonstrated in test cases involving large aggregates. Compared to existing methods, \ac{omer} shows greater accuracy, particularly in the range from visible to submillimetre wavelengths. The gain in computing time depends on the grain but can exceed two orders of magnitude.}
   {The proposed method provides a fast and accurate approximation for estimating the optical properties of materials with complex shapes and paves the way for efficient computation cross-sections for complex dust grain aggregates with thin core-mantle structures or hierarchical aggregates of large sizes.}


   \maketitle

\acresetall
\section{Introduction}
\label{introduction}

Dust grains are present in most astrophysical environments and interact significantly with light through absorption, scattering, and emission. Therefore, a detailed knowledge of the optical properties of dust is essential to understand the radiative phenomena occurring in the \ac{ism}, molecular clouds, protostellar nebulae, and protoplanetary discs.
The optical properties of dust, whether interstellar, interplanetary, or atmospheric, are highly dependent on the shape and distribution of materials within grains \citep{kohler_dust_2015,min_multiwavelength_2016,kimura_light_2016,doner_impact_2017, ysard_optical_2018,carpine_small_2025}. Aggregation, in particular, is a natural consequence of the growth of dust by collisions in any environment \citep{lorek_local_2018, ormel_dust_2009} and induces porosity and irregular structure, the effects of which on emissivity and scattering must be taken into account in dust modelling.

Several methods exist to compute the optical properties of dust grains. For simple spherical particles, Mie theory \citep{bohren_absorption_1983} provides a fast and accurate analytical solution. For more complex grains, with non-spherical shapes, made of several materials, including porosity, or with an aggregated structure, such an analytical computation cannot be performed, and numerical methods must be used instead. The \ac{tmm} method \citep{mishchenko_t-matrix_1996} and the \ac{dda} methods \citep{purcell_scattering_1973,draine_discrete-dipole_1988} typically provide accurate results for non-spherical particles, especially for aggregated grains. Both methods yield similar results, with a difference in the contact area between monomers \citep[see][for instance]{kohler_aggregate_2011}. The downside of numerical methods is the computing time required: the larger the grains, the more demanding in computing resources the calculations become. For instance, the \ac{dda} method allows freedom in the shape of the grains but requires the shape to be discretised into dipoles, the size of which is driven either by the smallest detail of the grain's complexity , or by the smallest wavelength computed. With a fixed dipole size, the number of dipoles increases cubically with grain size, as does the required computing time. \\
Several approximation methods have been developed to overcome limitations in both \ac{dda} and \ac{tmm}. In numerical methods \acs{emt} \citep[Effective medium theory;][]{garnett_xii_1904, bruggeman_berechnung_1935} can be useful for simplifying inhomogeneous spheres, but it remains restricted to thin mantles or small inclusions, with a limited refractive index contrast between the different materials \citep{mishchenko_applicability_2016}. The \ac{emt} has also been used to simulate the porosity of aggregated grains with spheroidal analogues \citep{draine_spheroidal_2025}, providing a correct level of accuracy for the size parameter $x = 2\pi a / \lambda $ such that $ 0.157\lesssim x \lesssim 15.7 $. Although in theory the validity of the method is limited by the size parameter, we specify that the given range is derived from the study's test on a single grain size $a_\eff = \qty{0.25}{\um}$, which gives the validity range $\qty{0.1}{\um}\lesssim\lambda\lesssim\qty{10}{\um}$, but does not ensure the size parameter limits for any grain sizes. In the domain of aggregated grains,  \ac{dhs} \citep{jones_modelling_1988,min_shape_2003} provides another approximation. Although this method is quite good at reproducing the opacities of compact aggregates \citep{min_multiwavelength_2016}, it does not closely reproduce the optical properties of fluffy aggregates outside of the Rayleigh domain (wavelength small compared to grain size) \citep{tazaki_light_2018}. The \ac{mft} method is also a good candidate, but remains limited when multiple scattering becomes important. \citet{tazaki_light_2018} introduced the \ac{mmf}, an accurate improvement on the \ac{mft}, which is accurate under the condition that the refractive index verifies $\lvert m-1\rvert<2$\footnote{This condition is usually verified at wavelengths $\lambda\lesssim\qty{10}{\um}$ for amorphous carbon and silicate grains, yet graphite or iron might not meet this criterion at small wavelengths.}, but appears to systematically underestimate extinction and scattering beyond this range. This method and the refinement introduced in \citet{botet_light_2025} are currently the best approximations for computing large fluffy aggregates. Lastly, the study from \citet{jones_spheroidal_2026} proposes an analytical work to approximate spheroidal core-mantle grains, but restricted to wavelengths in the Rayleigh domain of the grains.

This study concentrates on the simplification of a \ac{dda} computation of an aggregated grain with 'complex' monomers,  for example, a core-mantle or structured and/or aggregated monomer,  by reducing the number of dipoles in the monomer while maintaining the accuracy of our calculations.
We present \ac{omer}, a novel method for representing a structurally complex, mantle-covered, or aggregated dust grain with a simple homogeneous sphere computed through optimisation. This approach allows for fast calculation of the optical properties of dust, making it a valuable tool for computing the properties of large aggregated grains.

This paper is organised as follows. We present the technicalities of the method in Sect. \ref{sec:methods}. In Sect. \ref{sec:Results} we show the efficiency of the method in approximating several dust grains. We then derive two test cases in Sect. \ref{sec:applications} to demonstrate the use of \ac{omer} grains. In Sect. \ref{sec:stregthsAndCaveats} we discuss the strengths and caveats of the method. Finally, we summarise the conclusions of the study in Sect. \ref{sec:Conclusions}.

\section{Methods}
\label{sec:methods}

\subsection{Computation of optical properties of aggregated grains with \ac{dda}}

The scattering of light by dust particles can be predicted, to some extent, by the optical properties of the particle, extinction, absorption, scattering, and backscattering cross-sections $C_\ext, C_\abs, C_\sca$, $C_\back$, and the asymmetry parameter $g = <cos(\theta)>$ with $\theta$ the scattering angle. The complete description of the optical properties would be described by the Mueller matrix, but this study concentrates on the few key properties $C_\ext, C_\sca$, $C_\back$, and $g$ used in radiative transfer simulations to derive the photometry of emitted, scattered, and absorbed light, rather than trying to fit Mueller matrix elements, which can represent hundreds of parameters and can be more difficult to fit consistently.  To compute those properties for a set particle over a certain wavelength range, one can use the \ac{dda} method, which provides a precise numerical tool for the computation of arbitrary particle shapes and compositions. In particular, \ac{dda} is convenient for computing the optical properties of aggregated grains with complex structures and/or inhomogeneous compositions. The limit of this method lies in the size of the dipoles: either the smallest wavelength at which the computations are run, or, in our case, the smallest details of the structure that one wants to represent determine the size of the dipoles $d$, and the required computing time, which scales in $(a_\eff/d)^3$, can quickly become prohibitive. We note that the T-Matrix \citep{mishchenko_t-matrix_1996} method could outperform \ac{dda} in computational cost for such large aggregated grains. However, the T-Matrix method relies on the assumption that monomers are joined by single point contact. During aggregation processes, contact between monomers is expected to occur with a contact area \citep{dominik_physics_1997}, according to the Hertz contact model \citep{hertz_ueber_1882}, which models the physical interaction and interpenetration of two sticking grains. In this context, the assumption of point contact between monomers can lead to a significant underestimate of particle absorption \citep{kohler_aggregate_2011}. Because this aspect is particularly critical for the optical properties of aggregates in which many monomers are in contact, we chose to use the \ac{dda} method for our computations.

An aggregated grain (hereafter referred to as Agg.) is made up of a collection of monomers (hereafter referred to as Mon.), which are themselves grains with a more or less complex constitution. The simplest case would be homogeneous spherical monomers (Mon.sph.), but we can also consider core-mantle (two material) monomers (Mon.CM) or even monomers, which are themselves aggregates of small spheres (Mon.agg.). The respective aggregates are referred to as Agg.sph., Agg.CM, and Agg.hier. (for hierarchical aggregate). In the case of Agg.sph., the only geometric constraint on dipole size is the monomer size, and dipoles can be as large as the monomer size itself in some approximations \citep{okamoto_light_1998}. When computing Agg.CM, a minimum of three dipoles in the mantle thickness of the monomer is required for an accurate representation \citep[e.g.][]{kohler_dust_2015}. Lastly, when computing Agg.hier., the dipole size has to be determined by the smallest detail (monomer or mantle of the monomer) of the Mon.agg.. In the last two cases, the number of dipoles required to build large grains becomes computationally prohibitive for large aggregates. The aim of the OMER method, presented hereafter, is to simplify all Agg. to an Agg.sph. equivalent case.

\subsection{Implementation}
\label{subsec:implementation}

The OMER\footnote{The code for running the \ac{omer} will be made publicly available at publication of the paper.} method aims to simplify a complex Mon. grain, of which the optical properties $C_\ext, C_\sca, C_\back$, and $g$ are already known. This consists of building an 'equivalent homogeneous sphere', made of a virtual material, with a hypothetical refractive index $(n_\opt,k_\opt)$ that would closely reproduce the optical properties of the complex Mon. to then be able to create a large aggregate from this new OMER material. 

The size of the equivalent sphere $a_{\eff,\mathrm{OMER}}$ is defined by the apparent size of the real grain, in order to finely match the optical cross-sections. In the case of a spherical core-mantle grain, we take the grain size $a_\eff$, which is the grain external radius that includes both the core and the mantle. In the case of an aggregate, we take a surface equivalent radius $a_\mathrm{area} = \sqrt{{\langle 
\mathcal{A}\rangle }/{\pi}}$ as defined in \citet{kozasa_optical_1992}, where $\langle 
\mathcal{A}\rangle$ is the cross-sectional area of the grain averaged over three orthogonal directions.

In our method, finding $(n_\opt,k_\opt)$ is made possible by the fact that computing the optical properties of homogeneous and core-mantle spheres is fast and easy with the Lorenz-Mie theory. This method has been implemented in several computer codes, starting with Bohren and Huffman's \texttt{BHMIE} FORTRAN code \citep{bohren_absorption_1983}, up to the Python packages \texttt{PyMieScatt} \citep{sumlin_retrieving_2018} or \texttt{miepython} \citep{prahl_miepython_2025} that we use in this study. We define the function 
\begin{equation}
    \mathrm{Mie}_{\lambda,a_\eff}: \bigl(n(\lambda),k(\lambda)\bigr) \xrightarrow{} \bigl(C_\ext(\lambda), C_\sca(\lambda), C_\back(\lambda),g(\lambda)\bigr),
\end{equation}
which gives the optical properties of a homogeneous sphere of size $a_\eff$ and refractive index $(n,k)$ at wavelength $\lambda$, in other words, exactly what can be computed fast with \texttt{miepython}.

From this quick computation, we can perform an optimisation at a wavelength $\lambda$: we fix a search range for $(n_\opt(\lambda),k_\opt(\lambda))$ and perform a global optimisation of the function $\mathrm{Mie}_{\lambda,a_{\eff,\mathrm{OMER}}}$ on these two parameters, with the objective being the optical properties of the exact grain $C^{\obj}_\ext(\lambda), C^{\obj}_\sca(\lambda), C^{\obj}_\back(\lambda)$, and $g^{\obj}(\lambda)$. Optimisation is performed using the \texttt{shgo} function of the \texttt{scipy.optimize} Python package. This operation is repeated independently for every wavelength $\lambda$ in the desired range. The user can choose the minimisation function and the parameters fitted to their convenience and the specificities of their cases. In particular, we note that replacing $C_\ext$ by $C_\abs$ as a target parameter is a wise choice for accurate depiction of grains with $a_\eff \gg \lambda$. In the tests run in this paper, we use the equally-weighted sum of the squared normalised residuals 

\begin{multline}
    \left( \frac{C_\ext(\lambda) - C^{\obj}_\ext(\lambda)  }{C^{\obj}_\ext(\lambda)}\right)^2 + \left( \frac{C_\sca(\lambda) - C^{\obj}_\sca(\lambda)  }{C^{\obj}_\sca(\lambda)}\right)^2 \\ 
    \hfill+ \left( \frac{C_\back(\lambda) - C^{\obj}_\back(\lambda)  }{C^{\obj}_\back(\lambda)}\right)^2 +\left( \frac{g(\lambda) - g^{\obj}(\lambda)  }{g^{\obj}(\lambda)}\right)^2 ,
\end{multline}

\noindent to ensure that each parameter is fitted with the same importance as the others, regardless of their absolute value or unit.

We emphasise that the optimisation process to determine the optimal values of $n_{\opt}$ and $k_{\opt}$ for a given grain is performed independently for each wavelength. This approach not only saves time during the optimisation process, but also yields more accurate results. Notably, this process does not attempt to match the Kramers-Kronig relations. Therefore, the values obtained of $n_\opt$ and $k_\opt$ do not necessarily reflect the complex refractive index of a real material. In particular, there is no guarantee that the values of $n_\opt$ and $k_\opt$ will verify the Kramers–Kronig relations and they may exhibit unusual fluctuations. However, this does not affect the results of the purely numerical method of approximating large Agg. grains with \ac{omer} monomers as we show in the following. Furthermore, such fluctuations are mostly dependent on the size parameter $x = 2 \pi a / \lambda$, and smooth out when used in a dust population of different grain sizes.

\section{Results}
\label{sec:Results}

\subsection{Spherical core-mantle grains}
\label{subsec:CM}

\begin{figure*}[t]
\centering
\begin{subfigure}{.5\linewidth}
  \centering
  \subfigimg[width=\linewidth,clip]{pos=ur, hsep = \linewidth-2.5cm, vsep =1cm}{\small (a)}{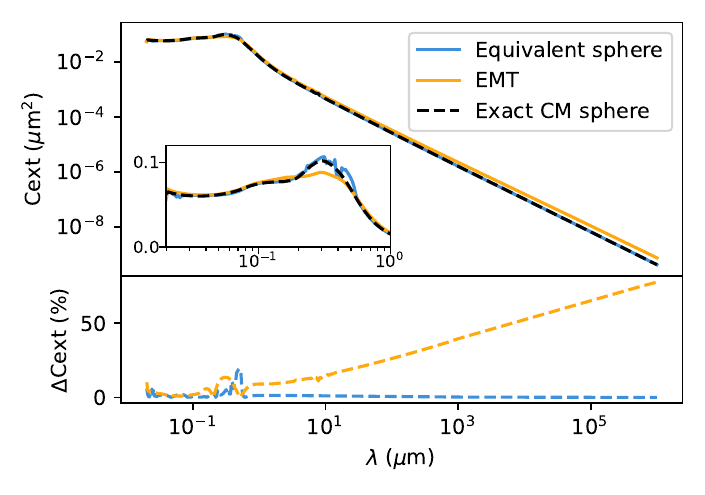}
  \phantomsubcaption
  \label{fig:aC(a)}
\end{subfigure}%
\begin{subfigure}{.5\linewidth}
  \centering
  \subfigimg[width=\linewidth,clip]{pos=ur, hsep = 0.7cm, vsep =1cm}{\small (b)}{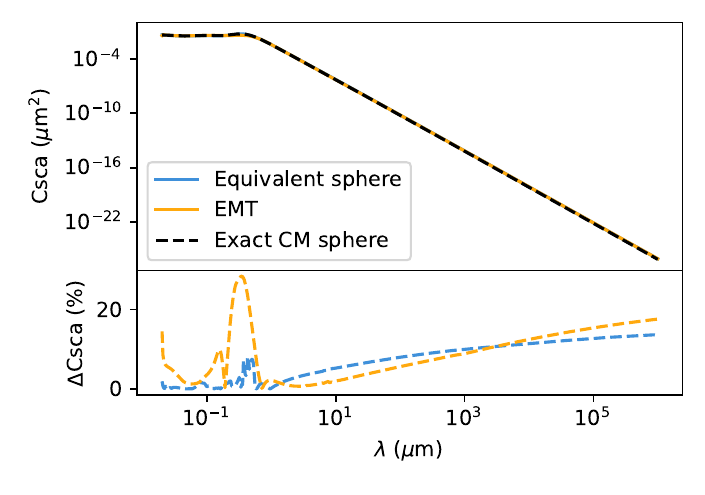}
  \phantomsubcaption
  \label{fig:aC(b)}
\end{subfigure}
\newline
\begin{subfigure}{.5\linewidth}
  \centering
  \subfigimg[width=\linewidth,clip]{pos=ur, hsep = 0.7cm, vsep =1cm}{\small (c)}{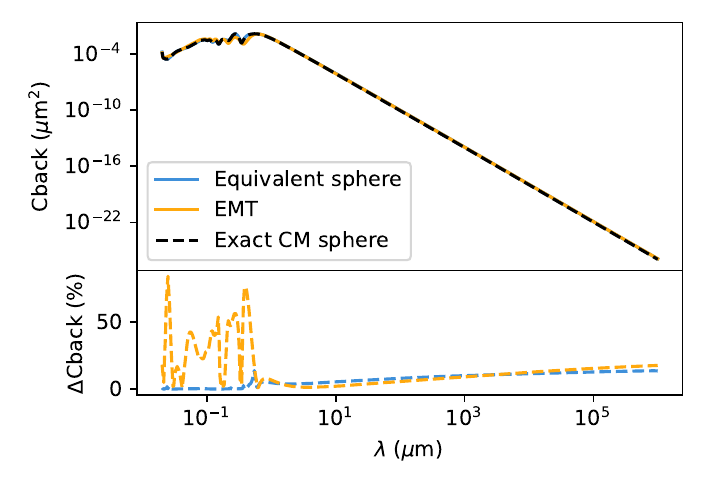}
  \phantomsubcaption
  \label{fig:aC(c)}
\end{subfigure}%
\begin{subfigure}{.5\linewidth}
  \centering
  \subfigimg[width=\linewidth,clip]{pos=ur, hsep = 0.7cm, vsep =1cm}{\small (d)}{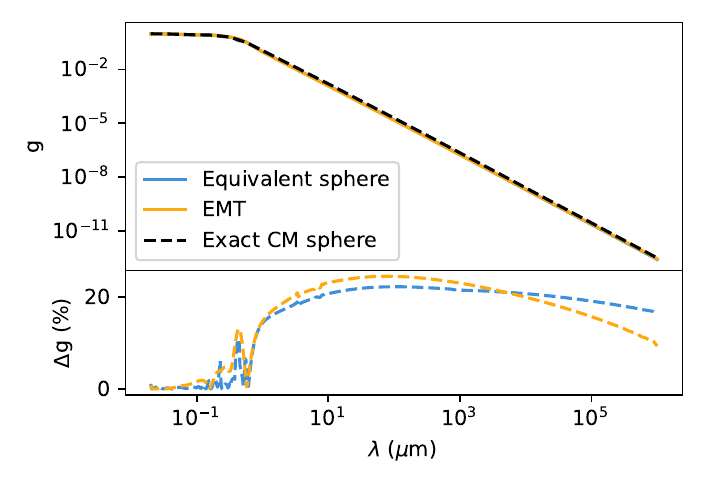}
  \phantomsubcaption
  \label{fig:aC(d)}
\end{subfigure}%
\caption{Optical properties of a \qty{100}{\nm} spherical core-mantle grain with an aCH core and a \qty{20}{\nm}-thick aC mantle (dashed black line), the equivalent representation of the same grain with OMER (blue) and the equivalent representation with \ac{emt} (yellow): extinction (a), scattering (b), backscattering (c) cross-sections, and asymmetry parameter (d). Bottom panels show the absolute value of the relative difference between the exact \texttt{miepython} computation and the equivalent representations. In panel (a), a zoom-in of the optical part of the curve is plotted with a linear y-scale.}
\label{fig:aC}
\end{figure*}

\begin{figure*}[h]
\centering
\begin{subfigure}{.5\textwidth}
  \centering
  \subfigimg[width=\linewidth]{pos=ur, hsep = 0.7cm, vsep =0.7cm}{\small (a)}{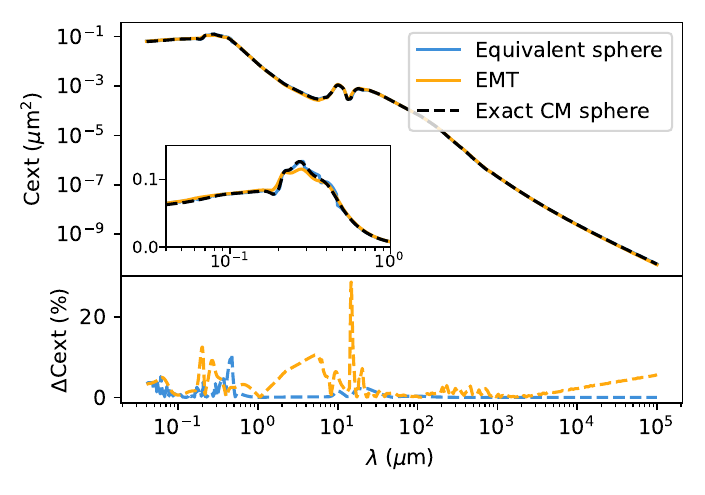}
  \phantomsubcaption
  \label{fig:aSil(a)}
\end{subfigure}%
\begin{subfigure}{.5\textwidth}
  \centering
  \subfigimg[width=\linewidth]{pos=ur, hsep = 0.7cm, vsep =0.7cm}{\small (b)}{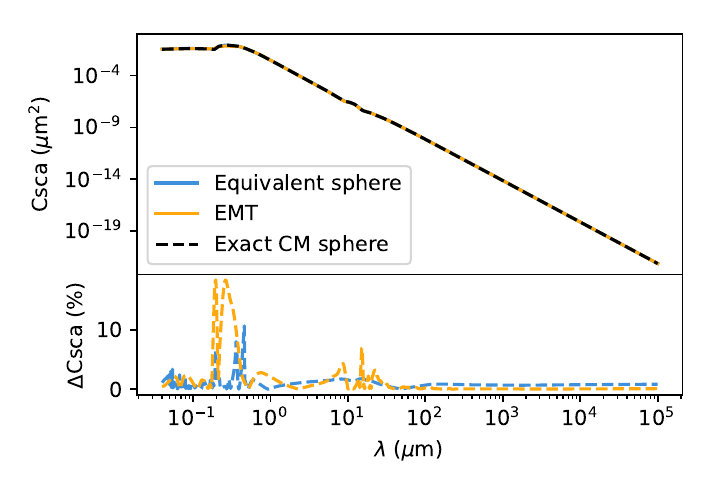}
  \phantomsubcaption
  \label{fig:aSil(b)}
\end{subfigure}\\%
\begin{subfigure}{.5\textwidth}
  \centering
  \subfigimg[width=\linewidth]{pos=ur, hsep = 0.7cm, vsep =0.7cm}{\small (c)}{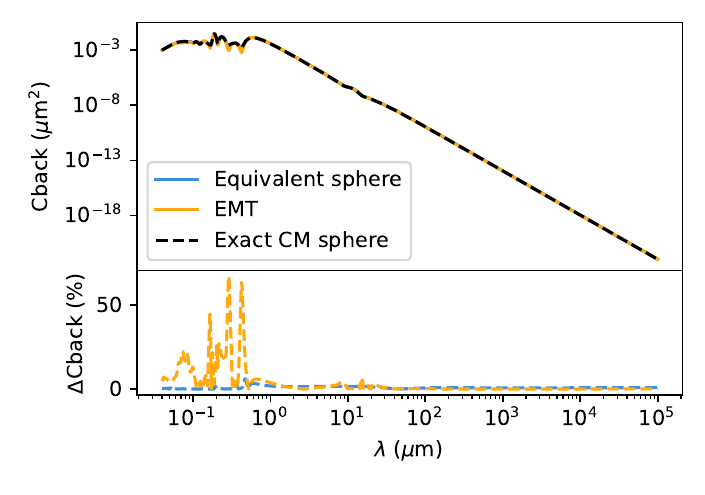}
  \phantomsubcaption
  \label{fig:aSil(c)}
\end{subfigure}%
\begin{subfigure}{.5\textwidth}
  \centering
  \subfigimg[width=\linewidth]{pos=ur, hsep = 0.7cm, vsep =0.7cm}{\small (d)}{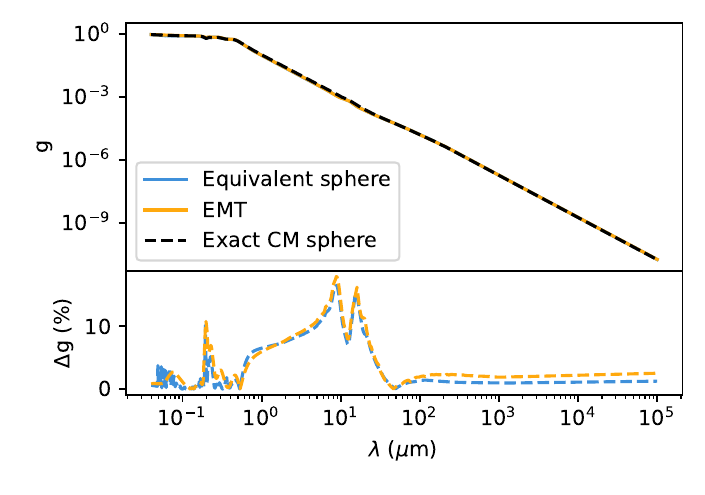}
  \phantomsubcaption
  \label{fig:aSil(d)}
\end{subfigure}%
\caption{Optical properties of a \qty{100}{\nm} spherical core-mantle grain with an aSil core and a \qty{5}{\nm}-thick aC mantle (dashed black line), the equivalent representation of the same grain with OMER (blue) and the equivalent representation with \ac{emt} (yellow): extinction (a), scattering (b), backscattering (c) cross-sections, and asymmetry parameter (d). Bottom panels show the absolute value of the relative difference between the exact \texttt{miepython} computation and the equivalent representations. In panel (a), a zoom-in of the optical part of the curve is plotted with a linear y-scale.}
\label{fig:aSil}
\end{figure*}

The first use of the method is the simplification of a core-mantle Mon. grain. This grain is composed of a spherical core with radius $a_\mathrm{core}$, layered with a shell of different material with thickness $t$, making the total grain a sphere of effective radius $a_\eff= a_\mathrm{core}+t$. In this section, we consider two Mon. grains with $a_\eff=\qty{100}{\nm}$, with the optical properties described in the THEMIS 2.0 model \citep{ysard_themis_2024}, namely one aSil/aC grain of amorphous silicate coated with a \qty{5}{\nm}-thick mantle of hydrogen-poor carbon and one aCH/aC grain of amorphous hydrogen-rich carbon coated with a \qty{20}{\nm}-thick mantle of hydrogen-poor carbon.

The challenge in this situation is the correct representation of the mantle in the grain when the grain is no longer isolated but forms part of a larger aggregate. For an accurate representation within the scope of \ac{dda}, at least three dipoles are required in the thickness of the mantle. In the case of a \qty{5}{\nm} mantle, nearly $10^6$ dipoles would be required to represent a \qty{100}{\nm} Mon. grain. When using such a grain in Agg. of several hundreds of monomers, the number of dipoles becomes too large to maintain a reasonable computation time with this representation. In addition, to model monomers of larger sizes, the dipole size remains determined by the mantle thickness,  resulting in a dramatic increase in the number of dipoles required. Using the OMER method, we can substitute Mon.CM. with a homogeneous \qty{100}{\nm} sphere of virtual material $(n_\opt,k_\opt)$. Using the OMER sphere instead of the real core-mantle grain lifts the computation limit put by the mantle thickness, allowing the size of the dipoles to be increased and the number of dipoles to be limited to only a few thousand.

The two test cases of THEMIS 2.0 aCH/aC and aSil/aC grains are presented in \cref{fig:aC} and \cref{fig:aSil}. In the case of spherical core-mantle grains, the Mie theory permits us to conduct the exact calculation of the optical properties of the grains to which we compare the OMER approximation. For further comparison, we also display the performance of another commonly used approximation method, the Bruggeman \ac{emt} \citep{bohren_absorption_1983}. This method was already used in the THEMIS 2.0 model for large aSil/aC grains, but not for aCH/aC grains due to the poor results obtained with these grains because of the high contrast between the imaginary parts of their optical constants at long wavelengths \citep{ysard_themis_2024}.

We can see in \cref{fig:aC} that we are able to reproduce the optical properties of the exact aCH/aC grain with an accuracy better than 15\% at all wavelengths for scattering and backscattering, and 20\% for the asymmetry parameter. Even more precisely, for wavelengths $\lambda \geq \qty{1}{\um}$, the extinction accuracy is better than 2\%, a better performance than that of the \ac{emt}. In the case of the aSil/aC grain, the results plotted in \cref{fig:aSil} show errors $\leq 10\%$ for all cross-sections, and $\leq 17\%$ for the asymmetry parameter from UV to microwave. Again, for wavelengths $\geq \qty{1}{\um}$, the error falls below 2\%, both for the scattering and extinction cross-sections in the aSil/aC case. The accuracy of \ac{omer} is comparable to that of  \ac{emt} for silicate grains at long wavelengths, but performs significantly better in the UV to NIR range. The same is true for carbonaceous grains. The \ac{omer} method shows better results for the aSil/aC grain here, whereas the properties are difficult to recover perfectly for the aCH/aC grain over the entire wavelength range. However, the clear superiority of \ac{omer} over \ac{emt} in the case of these spherical core-mantle grains should be noted.

Although this section focuses on \qty{100}{\nm} Mon. grains, we note that the approximation of the core-mantle spheres with \ac{omer} remains valid for different grain sizes, as described in \cref{app:size}. Moreover, it is also possible to use \ac{omer} spheres to approximate spheroidal core-mantle Mon. grains, as detailed in \cref{app:spheroid}. This proves the versatility of the method and its potential for creating aggregates of pseudo-spheroids.

\subsection{Aggregates}
\label{subsec:agg}

\begin{figure*}[h]
\centering
\begin{subfigure}{.5\textwidth}
  \centering
  \subfigimg[width=\linewidth]{pos=ur, hsep = \linewidth-2.5cm, vsep =1cm}{\small (a)}{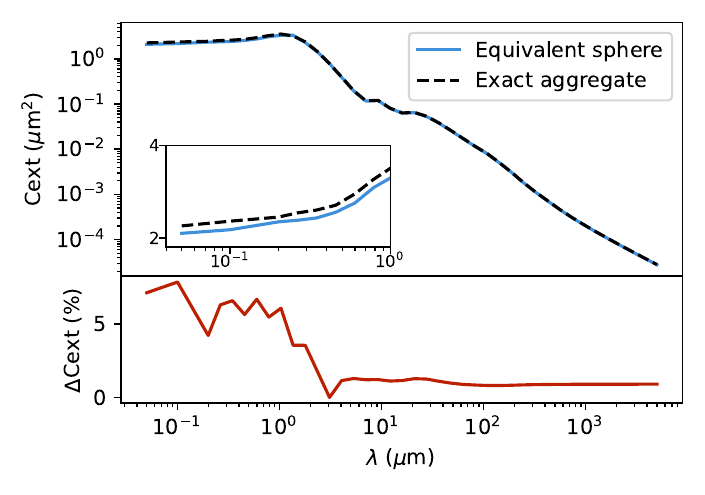}
  \phantomsubcaption
  \label{fig:Agg(a)}
\end{subfigure}%
\begin{subfigure}{.5\textwidth}
  \centering
  \subfigimg[width=\linewidth]{pos=ur, hsep = 0.7cm, vsep =1cm}{\small (b)}{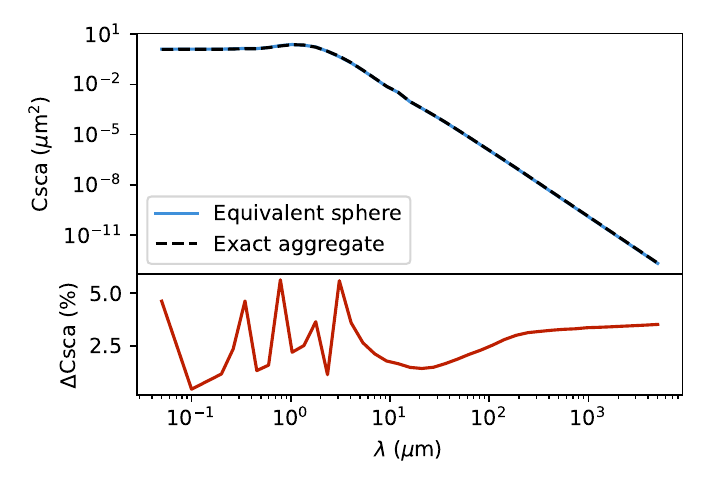}
  \phantomsubcaption
  \label{fig:Agg(b)}
\end{subfigure}\\%
\begin{subfigure}{.5\textwidth}
  \centering
  \subfigimg[width=\linewidth]{pos=ur, hsep = 0.7cm, vsep =1cm}{\small (c)}{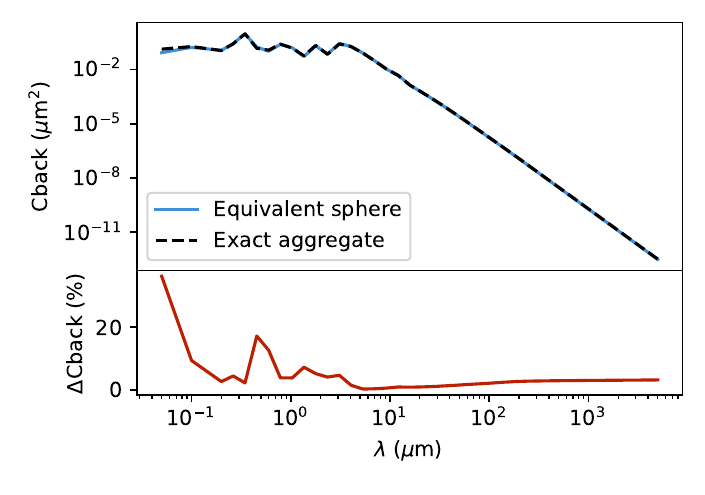}
  \phantomsubcaption
  \label{fig:Agg(c)}
\end{subfigure}%
\begin{subfigure}{.5\textwidth}
  \centering
  \subfigimg[width=\linewidth]{pos=ur, hsep = \linewidth-2.5cm, vsep =1cm}{\small (d)}{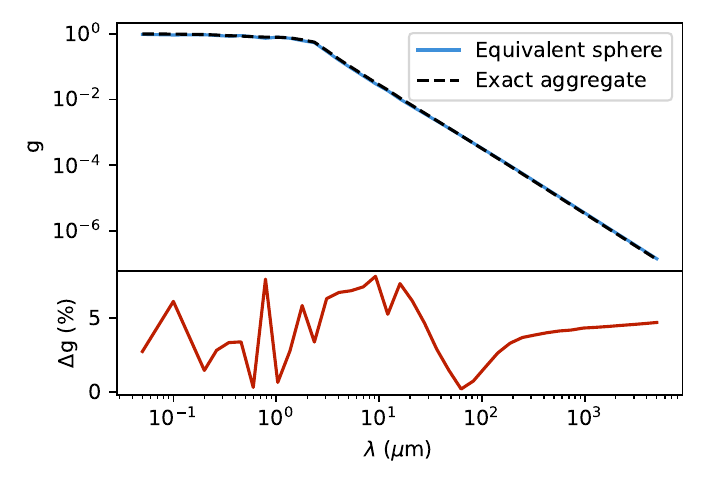}
  \phantomsubcaption
  \label{fig:Agg(d)}
\end{subfigure}%
\caption{Optical properties of a \qty{489}{nm} aggregate grain made of aCH/aC and aSil/aC core-mantle \qty{100}{\nm} spherical monomers (dashed black line) and the equivalent representation of the same grain with OMER (blue), as described in Sect. \ref{subsec:agg}: extinction (a), scattering (b), backscattering (c) cross-sections, and asymmetry parameter (d). Bottom panels represent the absolute value of the relative difference between the exact \ac{dda} computation and the OMER equivalent representation. In panel (a), a zoom-in of the optical part of the curve is plotted with linear y-scale.}
\label{fig:Agg}
\end{figure*}

The second test case of the OMER method is run on a Mon.agg. aggregated grain, of effective radius $a_\eff=\qty{489}{nm}$. The simplification of this grain will be useful for the building of hierarchical aggregates. The aggregate described here is made of \qty{100}{nm} aSil/aC and aCH/aC core-mantle THEMIS 2.0 spheres, the former represented by homogeneous \ac{emt} material, and the latter represented with a proper core-mantle structure. Therefore, the dipole size is constrained by the thickness of the aC mantle on top of the carbonaceous monomers. Even without a mantle, an arbitrary aggregate would require a greater number of dipoles than a sphere to accurately represent the details of its substructure. The OMER equivalent will lift these constraints and permit large Agg.hier. made of small Mon.agg. to be built, by simplifying the latter to homogeneous spheres. This limits the number of dipoles required in the whole grain, thus reducing the computation time.

The process of finding the OMER representation of this grain is similar to the one described in Sect. \ref{subsec:CM}, which means that an exact initial calculation must be performed. In the case of an aggregate, Mie theory obviously cannot be used, so we use \ac{dda} thanks to the \texttt{ADDA} routine from \citet{yurkin_discrete-dipole-approximation_2011}. The equivalent sphere for this Mon.agg. needs to reproduce the properties of the Mon.agg., and especially its cross-sectional properties, which can only be reproduced if the geometrical properties of both grains are the same. For this reason, instead of keeping the same effective radius $a_\eff$, which in this case would be the equivalent volume radius, we use the surface-equivalent radius $a_\mathrm{area}  = \sqrt{{\langle \mathcal{A}\rangle }/{\pi}}$ defined in Sect. \ref{subsec:implementation}, to ensure identical geometric cross-sections between the exact grain and its OMER equivalent. A Mon.agg with $a_\eff=\qty{489}{nm}$ has $a_\mathrm{area} = \qty{559}{nm}$.

The results are shown in \cref{fig:Agg}: $C_\ext, C_\sca$, and $g$ are retrieved with an accuracy better than 8\% at all wavelengths. The backscattering has a slightly higher error compared to the other quantities at short wavelengths, with errors up to $35\%$ in UV, but the errors drop to $\leq 17 \%$ for $\lambda > \qty{0.1}{\um}$ and $\leq 7 \%$ for $\lambda > \qty{1}{\um}$. The OMER method is therefore effective in the case of core-mantle monomer aggregates.

\section{Implementation in practical cases}
\label{sec:applications}

This section examines the accuracy of OMER in two practical cases. The challenging aspect lies in computing the reference case, as the OMER method is designed to simplify grains that would otherwise be too computationally intensive to run effectively. Therefore, we performed our tests on two simple cases. First, we consider an aggregate composed of only two aCH/aC spheres (see \cref{fig:2sph3D(a)}), over the entire wavelength range from \qty{50}{nm} to \qty{5}{mm}. Secondly, we study a 'full-sized' Agg.hier grain (see \cref{fig:Agghier3D(a)}), but because of the extensive need of computational resources, we only derive the properties at six wavelengths: \qty{0.45}{\um}, \qty{5.34}{\um}, \qty{20.98}{\um}, \qty{323.85}{\um},  \qty{2.89}{mm}, and \qty{3.80}{mm}.

\subsection{Two-sphere aggregate}

\begin{figure*}[h]
\centering
\begin{subfigure}{.18\textwidth}
  \centering
  \subfigimg[trim={55cm 0 0 0},width=\linewidth,clip]{pos=ur, hsep = 0.5cm, vsep =3cm}{\small (a)}{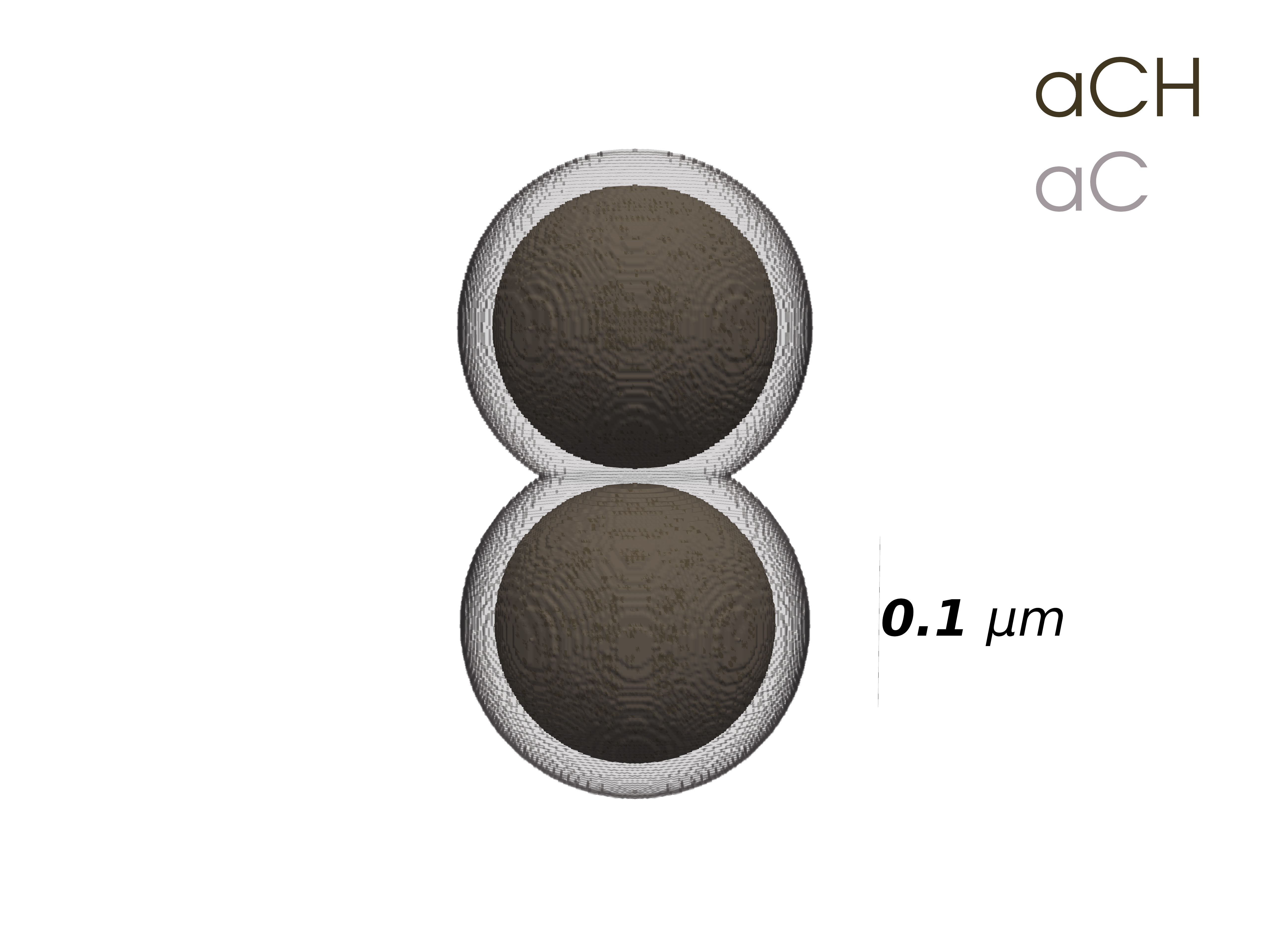}
  \phantomsubcaption
  \label{fig:2sph3D(a)}
\end{subfigure}%
\begin{subfigure}{.18\textwidth}
  \centering
  \subfigimg[trim={55cm 0 0 0},width=\linewidth,clip]{pos=ur, hsep = 0.5cm, vsep =3cm}{\small (b)}{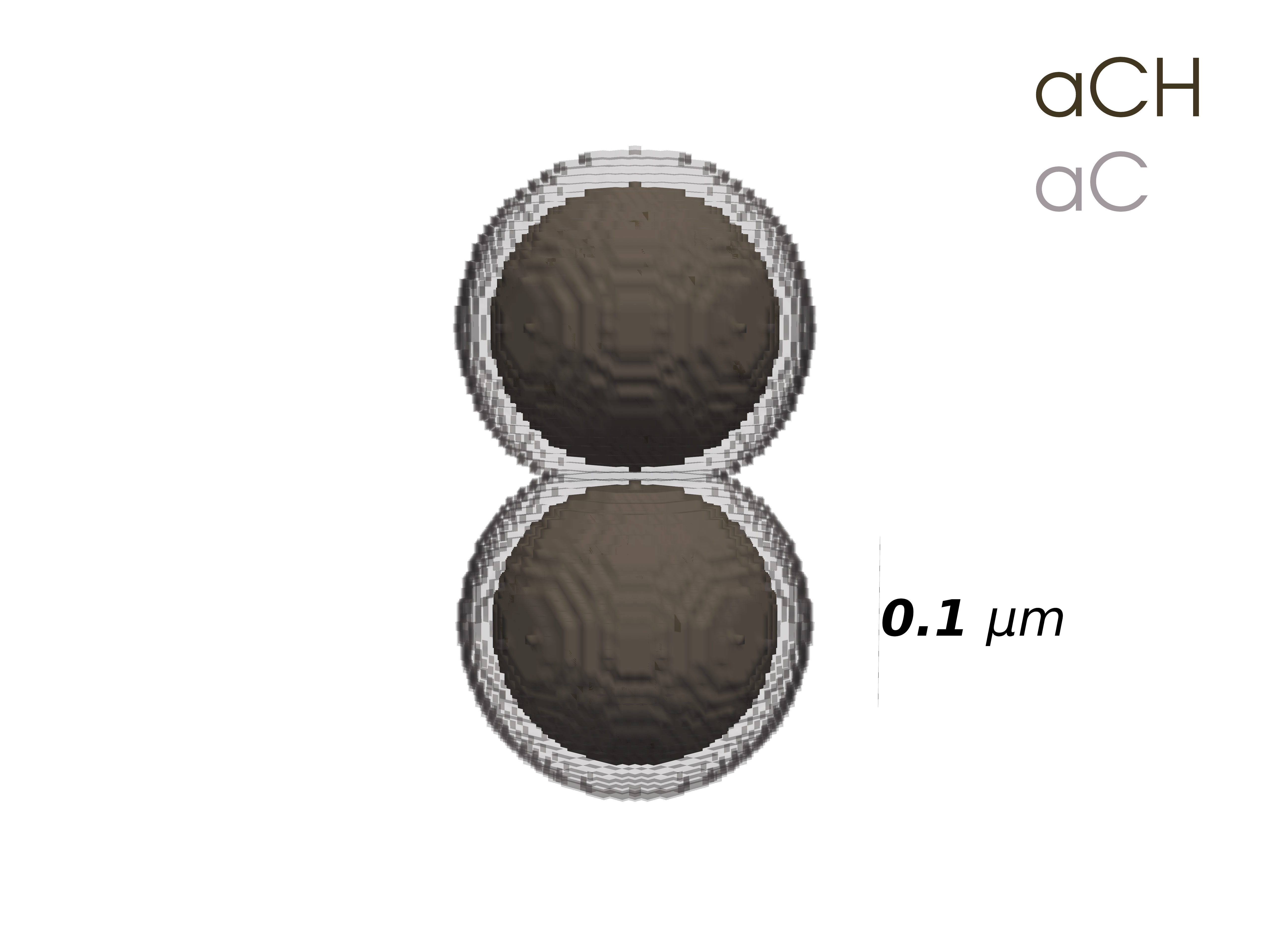}
  \phantomsubcaption
  \label{fig:2sph3D(b)}
\end{subfigure}%
\begin{subfigure}{.18\textwidth}
  \centering
  \subfigimg[trim={55cm 0 0 0},width=\linewidth,clip]{pos=ur, hsep = 0.5cm, vsep =3cm}{\small (c)}{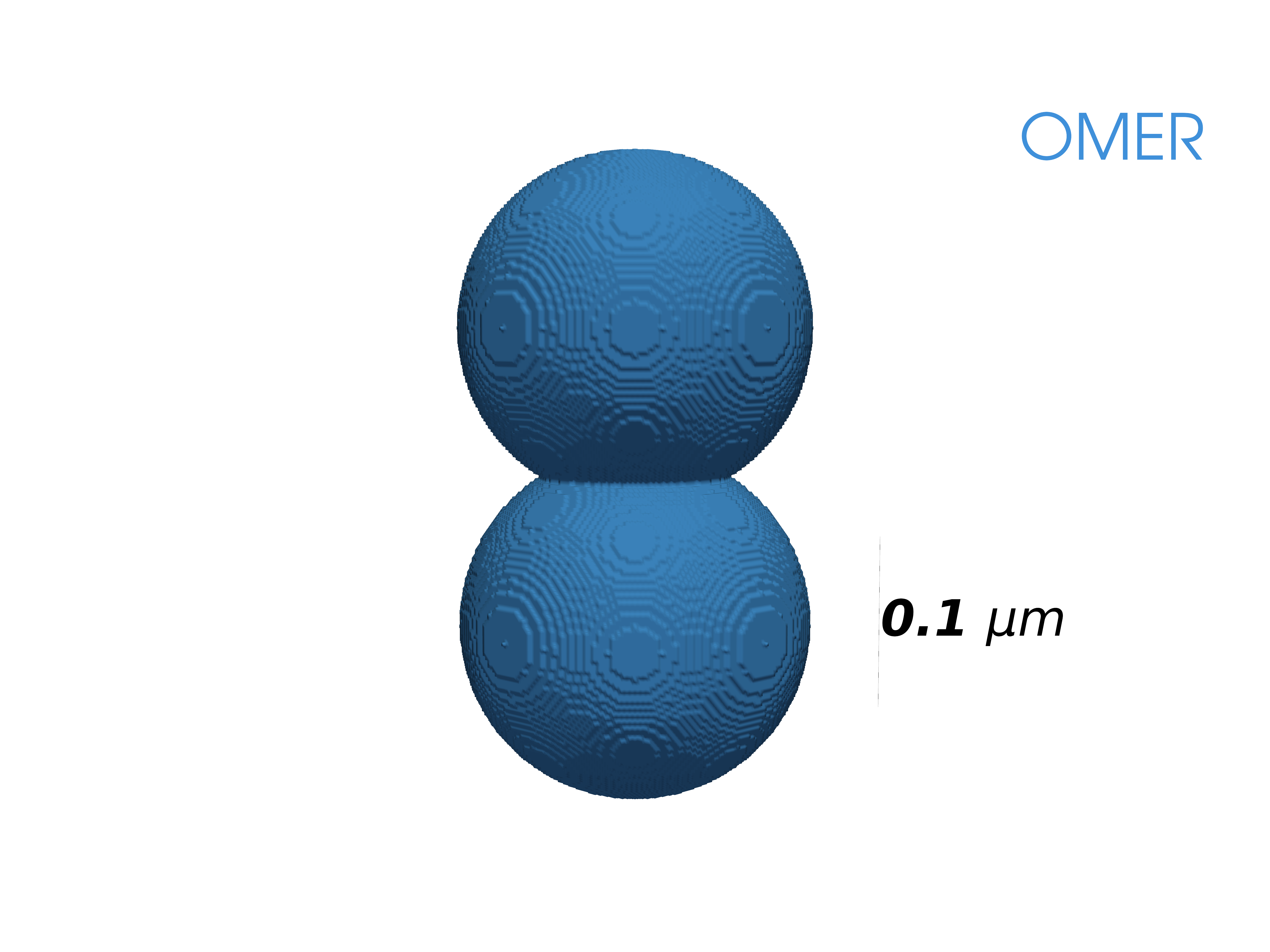}
  \phantomsubcaption
  \label{fig:2sph3D(c)}
\end{subfigure}%
\begin{subfigure}{.18\textwidth}
  \centering
  \subfigimg[trim={55cm 0 0 0},width=\linewidth,clip]{pos=ur, hsep = 0.5cm, vsep =3cm}{\small (d)}{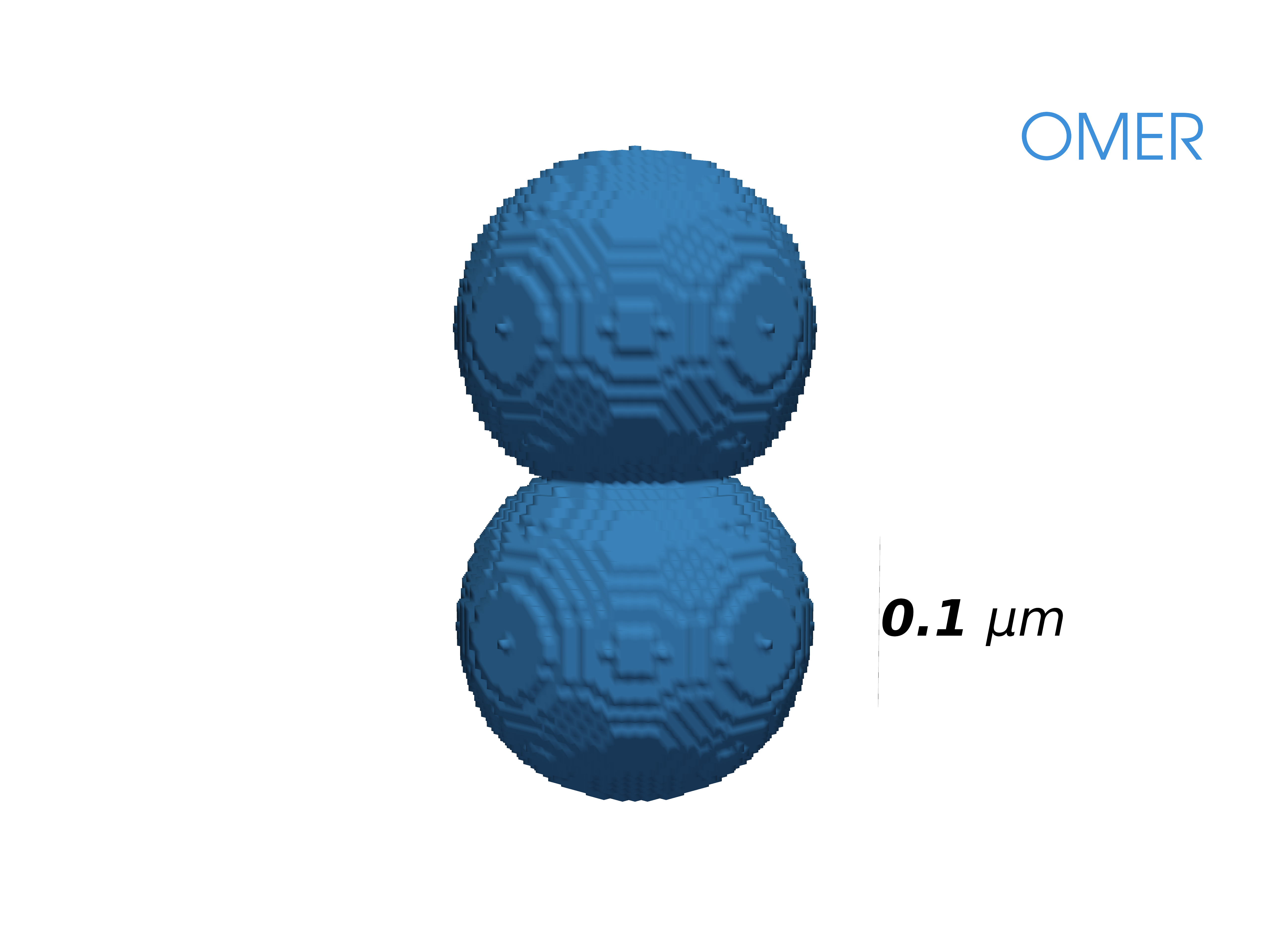}
  \phantomsubcaption
  \label{fig:2sph3D(d)}
\end{subfigure}%
\begin{subfigure}{.18\textwidth}
  \centering
  \subfigimg[trim={55cm 0 0 0},width=\linewidth,clip]{pos=ur, hsep = 0.5cm, vsep =3cm}{\small (e)}{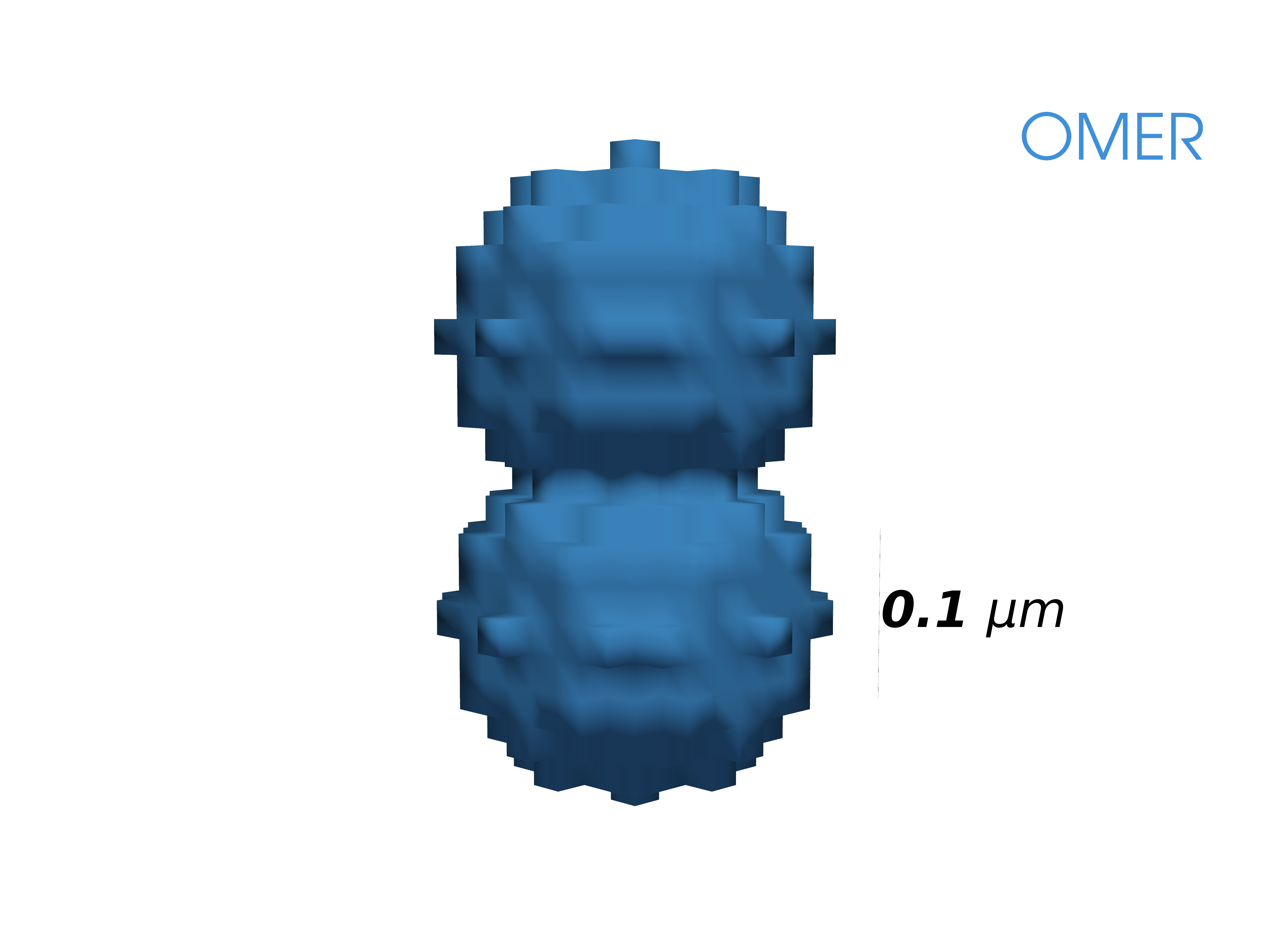}
  \phantomsubcaption
  \label{fig:2sph3D(e)}
\end{subfigure}%
\caption{3D representation of aggregate grains made of two \qty{100}{\nm} core-mantle spheres with aCH cores and \qty{20}{\nm}-thick aC mantles (a, b) and the equivalent representation of the same grain with OMER spheres (c, d, e), with different dipole sizes: $d=\qty{2}{nm}$ (a, c), $d=\qty{5}{nm}$ (b, d), and $d=\qty{20}{nm}$ (e).}
\label{fig:2sph3D}
\end{figure*}

As this practical case can be quickly calculated with \texttt{ADDA}, we derive the optical properties of the same grain, with different sets of parameters, as shown in \cref{fig:2sph3D}. The grain comprises two \qty{100}{nm} aCH/aC spheres featuring a \qty{20}{nm}-thick mantle, as described in Sect. \ref{subsec:CM}, with centres \qty{170}{nm} apart. The grain is computed with the exact core-mantle structure and materials on the one hand (\cref{fig:2sph3D(a),fig:2sph3D(b)}) and with the OMER material on the other hand (\cref{fig:2sph3D(c),fig:2sph3D(d),fig:2sph3D(e)}). As the OMER method aims to increase the dipole size $d$ in order to decrease the computation time, we iterate on $d$, setting it at $d =$ \qty{2}{nm} and \qty{5}{nm} for both representations, and an additional $d=\qty{20}{nm}$ for the OMER grain.

All results for the extinction and scattering cross-sections, as well as the asymmetry parameter ($C_\ext$, $C_\sca$, and $g$), are plotted in \cref{fig:2sph}. In the error plots, the exact \ac{dda} case with $d=\qty{2}{nm}$ is taken as the reference case. In \cref{fig:2sph(d)}, we also show the computing time required by \texttt{ADDA} to compute the optical properties of the grain at each wavelength and for one angle orientation.

We see that the errors in this case remain moderate, except for a peak around \qty{400}{nm}. For extinction, the OMER method provides an approximation of within $10\%$ with $d=\qty{20}{nm}$ and within $4\%$ with $d=\qty{5}{nm}$, which is even better than the exact \ac{dda} result for the same size of dipole. For scattering and asymmetry, the error remains within $20\%$ with $d=\qty{20}{nm}$, and within $15\%$ with $d=\qty{20}{nm}$. The only issue with this approximation is the peak in error at $\lambda =\qty{400}{nm} =4 a_0$, likely due to some scattering effects of the $a_0$-sized monomers. This peak reaches $40\%$ in the $C_\ext$ and $C_\sca$ errors, $75\%$ in $g$ with $d=\qty{20}{nm}$, and $40\%$ in $g$ with $d=\qty{20}{nm}$. In this case, the OMER method provides a fairly good approximation, giving results superior to those obtained by modelling the 'exact' mantle when the dipole size becomes comparable to the mantle thickness. However, the peak error due to the scattering effect should be kept in mind. In particular, the resulting extinction cross-section of the grain is fairly altered as seen \cref{fig:2sph(a)}, and what could be seen as an artificial spectral feature arises. We insist that this discontinuity is neither a molecular nor a crystalline feature, but an artificial discontinuity of the method. We show in \cref{app:crystal} that the OMER method does not have issues modelling fine crystalline features. Furthermore, we want to stress that the wavelength at which the discontinuity arises is linked to the monomer size: as pointed out in \cref{app:size}, the error peak in the OMER approximation predictably relates to the grain size as $\lambda=4a_0$. The discontinuity seen here at \qty{400}{\nm} with \qty{100}{\nm} monomers is expected to arise at \qty{800}{\nm} if \qty{200}{\nm} are used. We therefore insist that the spectral discontinuity that is created by the method in this single dust grain will be smoothed out when such a grain is incorporated into a dust population containing a whole range of grain sizes, whose optical properties will average.

The most striking aspect is the decrease in computing time allowed by the OMER method as the dipole size increases. The computing time is expected to scale roughly with the number of dipoles, $n_d$, so with a fixed grain size, it would scale as $d^{-3}$. Specifically, the total computation times are \qty{14}{\hcpu} and \qty{0.34}{\hcpu} for the exact grains, and \qty{7}{\hcpu}, \qty{0.18}{\hcpu}, and \qty{0.037}{\hcpu} for the OMER grains, with dipole sizes of \qty{2}{\nm}, \qty{5}{\nm}, and \qty{20}{\nm}, respectively. We also see in \cref{fig:2sph(d)} that the computing time diminishes as the dipole size increases from \qty{2}{\nm} to \qty{5}{\nm} and \qty{20}{\nm}, reducing respectively by 75 and 370 the computing time of the grain optical properties. In this case, the number of dipoles is large compared to the radius of the monomer, so we do not increase it further. However, we stress that in cases where the monomer is larger in size, with the same \qty{20}{nm} mantle, the size of the dipole would remain restricted to a few manometers for the exact DDA computation of an aggregate with core-mantle monomers. The number of dipoles would increase significantly, whereas with the OMER method, we could keep the size of the dipole constant relative to the size of the monomer and maintain a reasonable number of dipoles and computing times.

\begin{figure*}[h]
\centering
\begin{subfigure}{.5\textwidth}
  \centering
  \subfigimg[width=\linewidth]{pos=ur, hsep = \linewidth-2.5cm, vsep =1cm}{\small (a)}{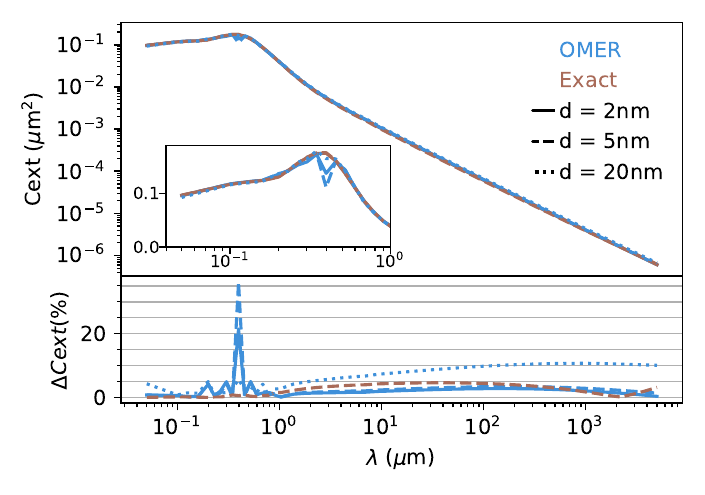}
  \phantomsubcaption
  \label{fig:2sph(a)}
\end{subfigure}%
\begin{subfigure}{.5\textwidth}
  \centering
  \subfigimg[width=\linewidth]{pos=ur, hsep = 1cm, vsep =1cm}{\small (b)}{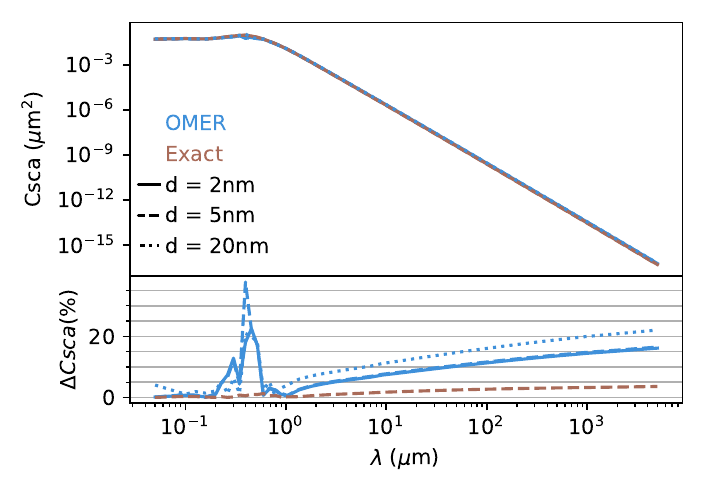}
  \phantomsubcaption
  \label{fig:2sph(b)}
\end{subfigure}\\%
\begin{subfigure}{.5\textwidth}
  \centering
  \subfigimg[width=\linewidth]{pos=ur, hsep = \linewidth-2.5cm, vsep =1cm}{\small (c)}{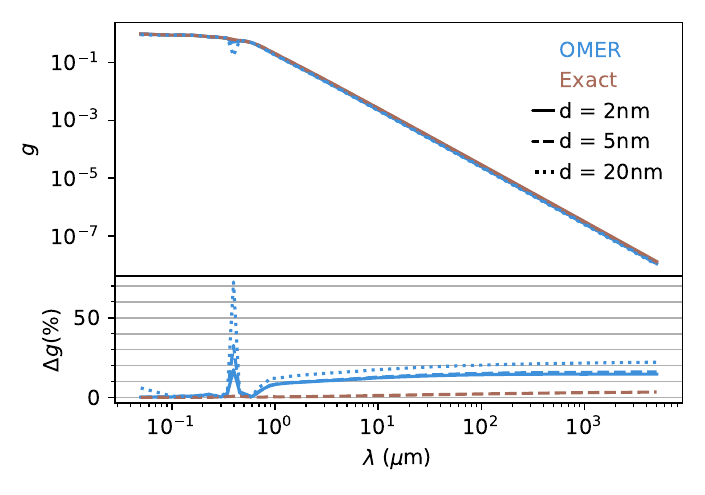}
  \phantomsubcaption
  \label{fig:2sph(c)}
\end{subfigure}%
\begin{subfigure}{.5\textwidth}
  \centering
  \subfigimg[width=\linewidth]{pos=ur, hsep = 1cm, vsep =1cm}{\small (d)}{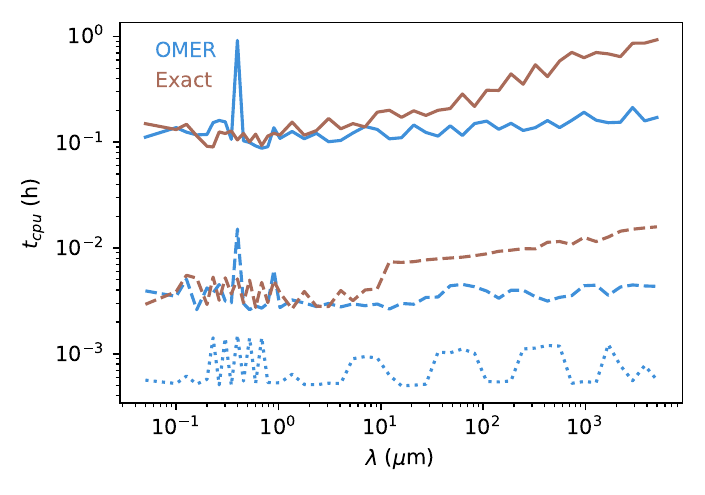}
  \phantomsubcaption
  \label{fig:2sph(d)}
\end{subfigure}%
\caption{Optical properties of an aggregate composed of two 100nm spherical monomers with aCH cores and 20 nm-thick aC mantles as represented in \cref{fig:2sph3D(a)} (brown) and of the equivalent representation of the same grain with
OMER spheres for monomers as represented in \cref{fig:2sph3D(c)} (blue): extinction (a), scattering (b) cross-sections, asymmetry parameter (c), and required CPU computing time (d) per orientation. Computations are made with several dipole sizes $d=\qty{2}{nm}$ (solid lines), $d=\qty{5}{nm}$ (dashed), and  $d=\qty{20}{nm}$ (dotted). Bottom panels in figures (a, b, c) show the absolute value of the relative difference between the exact \ac{dda} computation (core-mantle with $d=\qty{2}{nm}$) and the other representations. In panel (a), a zoom-in of the optical part of the curve is plotted with linear y-scale.}
\label{fig:2sph}
\end{figure*}

\subsection{Hierarchical aggregate}

For the second practical case, and maybe most relevant, we study the building of an Agg.hier using the OMER method to simplify the Mon.agg. This case corresponds to multiple-level aggregation, where aggregates coagulate among themselves to form what we call a 'cluster' of aggregates, or a hierarchical aggregate. In this case, we use \qty{489}{nm} Mon.agg. constituted of \qty{100}{nm} aSil/aC and aCH/aC spheres, just as in Sect. \ref{subsec:agg}, coagulated in a $\sim \qty{1.174}{\um}$ Agg.hier., as represented in \cref{fig:Agghier3D(a)}. 

In \cref{fig:Agghier}, we plot the comparison of the optical properties of the exact grain and the OMER approximate grain, with $C_\ext$, $C_\sca$, and $g$, along with the computational time comparison. The plotted time is the time required to compute, at each wavelength, the optical properties of the grain in one orientation. For a complete orientational averaging of an aggregate grain, a total of 24 orientations is required. For example, computing the OMER grain at all 40 wavelengths requires a total of  \qty{95}{\kilo\hcpu}. Using the OMER approximation in this case results in a significant time gain: the computation at the two largest wavelengths of the exact case requires \qty{3371}{\hcpu}, whereas at the same wavelengths the OMER case requires \qty{44}{\hcpu}. There is a difference of two orders of magnitude, even though the number of dipoles is only three times smaller (63M versus 188M) with the OMER method. This is likely due to the cost of parallelisation, as a simulation with these many dipoles requires a large number of CPUs, and it illustrates the difficulty of running grains with a really large number of dipoles. At the six wavelengths computed for the exact grain, the optical properties closely match the OMER approximation. Should computing times per dipole be similar between the exact grain and the OMER grain (which is the case at smaller wavelengths), the OMER method permits reducing the number of dipoles up to several orders of magnitude, providing much reduced required computing time in any case. Scattering is approximated with an error of less than 8.1\%, the asymmetry parameter with a small error of less than 2.5\%, and the extinction with an error of less than 15.7\%. It is interesting to note that the logarithmic slope of the extinction $\beta= - \d \log(C_\ext) / \d \log(\lambda)$ is almost perfectly approximated by the OMER method, with a difference of less than 0.2\% compared to the reference case.

Here, the OMER method proves to be a highly valuable tool. It allows us to approximate, with good accuracy, grains that are beyond the scope of reasonable computation because of their hierarchical structure by modelling them as simple spheres in \ac{dda} computations with a reduced number of dipoles. This monomer approximation can be extended to basically any grain for which the optical properties have already been computed, and this grain can then be used to build new, larger aggregates. For instance, we could use the Agg.hier. that we computed here with the OMER method as monomers to build third-level aggregates and extend computing capabilities to ever larger grains.

\begin{figure*}[h]
\sidecaption
\begin{subfigure}{6cm}
  \centering
  \subfigimg[trim={0 0 0 0},width=\linewidth,clip]{pos=ur, hsep = 0.5cm, vsep =3.5cm}{\small (a)}{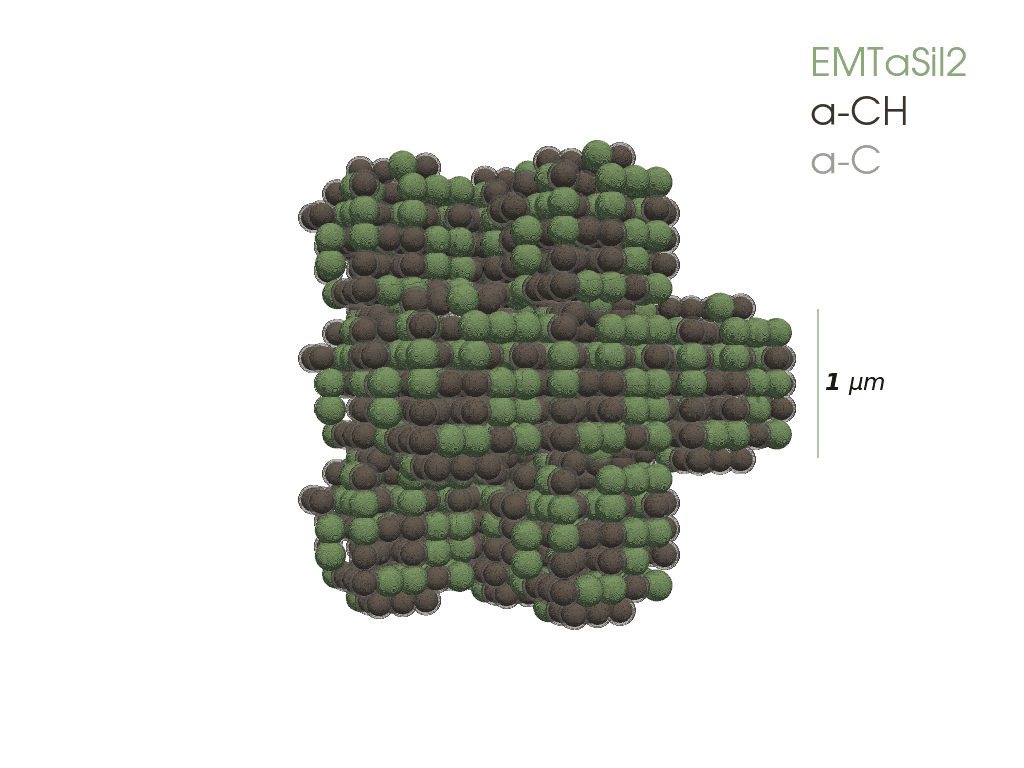}
  \phantomsubcaption
  \label{fig:Agghier3D(a)}
\end{subfigure}%
\begin{subfigure}{6cm}
  \centering
  \subfigimg[trim={0 0 0 0},width=\linewidth,clip]{pos=ur, hsep = 0.5cm, vsep =3.5cm}{\small (b)}{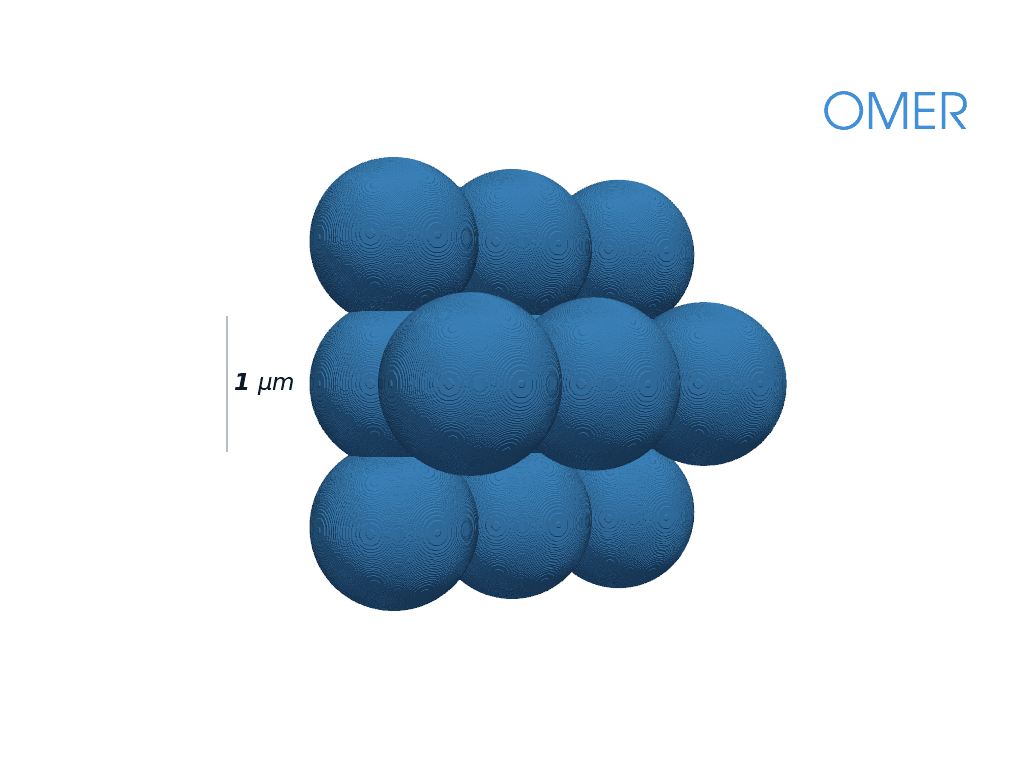}
  \phantomsubcaption
  \label{fig:Agghier3D(b)}
\end{subfigure}%
\addtocounter{figure}{-1}
\caption{3D representation of \qty{1174}{\nm} Agg.hier. made of \qty{489}{\nm} Mon.Agg aCH/aC and aSil/aC core-mantle monomers (a) and the equivalent representation of the same grain with OMER spheres for monomers (b).}
\label{fig:Agghier3D}
\end{figure*}

\begin{figure*}[h]
\centering
\begin{subfigure}{.5\textwidth}
  \centering
  \subfigimg[width=\linewidth]{pos=ur, hsep = \linewidth-2.5cm, vsep =1cm}{\small (a)}{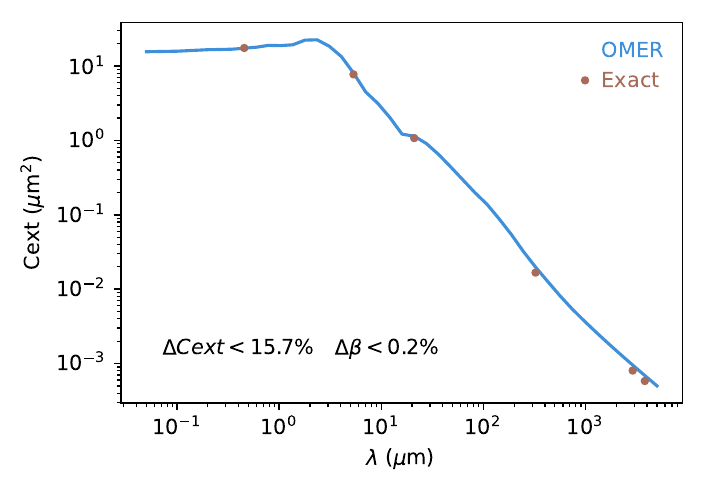}
  \phantomsubcaption
  \label{fig:Agghier(a)}
\end{subfigure}%
\begin{subfigure}{.5\textwidth}
  \centering
  \subfigimg[width=\linewidth]{pos=ur, hsep = \linewidth-2.5cm, vsep =1cm}{\small (b)}{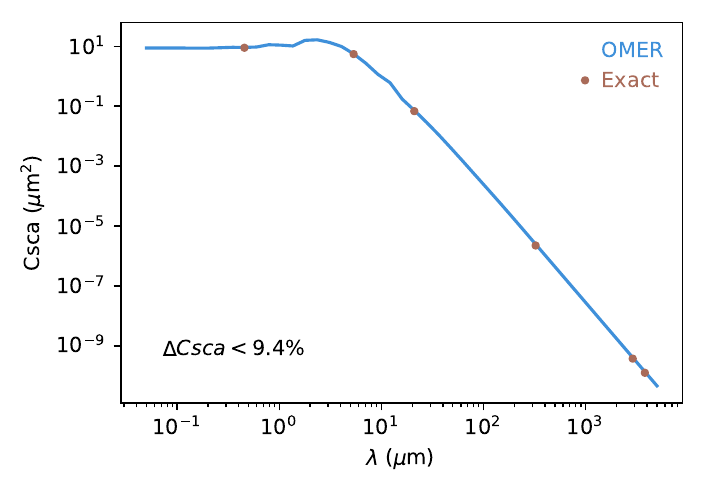}
  \phantomsubcaption
  \label{fig:Agghier(b)}
\end{subfigure}\\%
\begin{subfigure}{.5\textwidth}
  \centering
  \subfigimg[width=\linewidth]{pos=ur, hsep = \linewidth-2.5cm, vsep =1cm}{\small (c)}{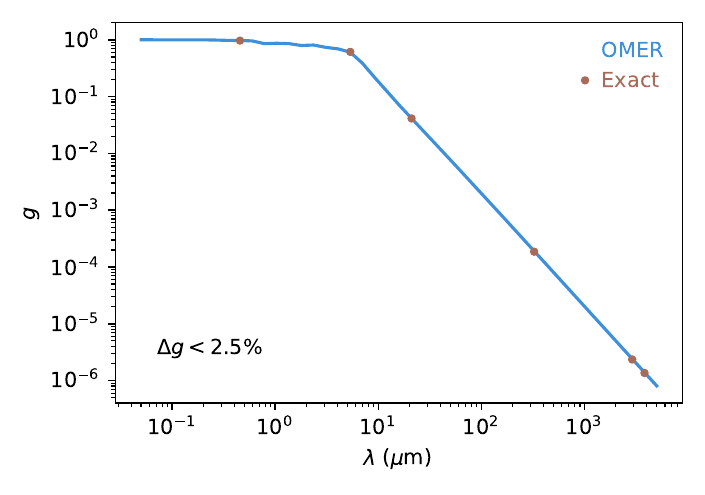}
  \phantomsubcaption
  \label{fig:Agghier(c)}
\end{subfigure}%
\begin{subfigure}{.5\textwidth}
  \centering
  \subfigimg[width=\linewidth]{pos=ur, hsep = \linewidth-2.5cm, vsep =0.7cm}{\small (d)}{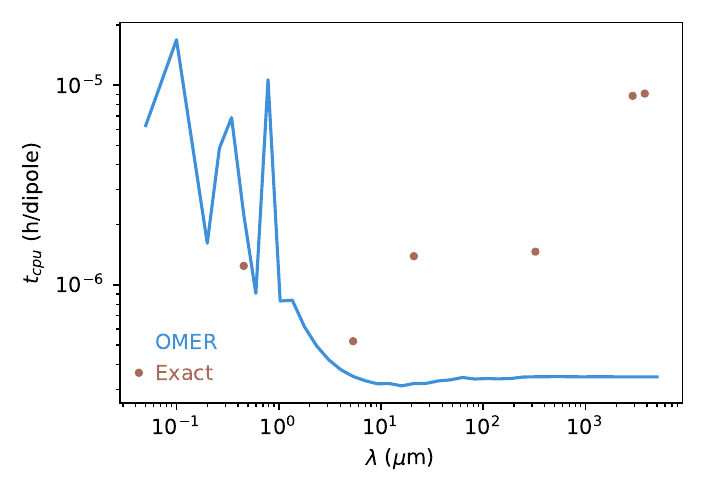}
  \phantomsubcaption
  \label{fig:Agghier(d)}
\end{subfigure}%
\caption{Optical properties of a \qty{1174}{\nm} Agg.hier. made of \qty{489}{\nm} Mon.Agg aCH/aC and aSil/aC core-mantle monomers as represented in \cref{fig:Agghier3D(a)} (brown) and the equivalent representation of the same grain with OMER spheres for monomers as represented in \cref{fig:Agghier3D(b)} (blue): extinction (a), scattering (b), cross-sections, asymmetry parameter (c), and required CPU computing time (d) per orientation and per dipole (24 orientations are needed to compute the whole grain angular averaging). Bottom panels in figures (a, b, c) show the absolute value of the relative difference between the exact \ac{dda} computation and the equivalent representations.}
\label{fig:Agghier}
\end{figure*}

\section{Strengths and caveats}
\label{sec:stregthsAndCaveats}

In this section we review the method that we introduced within the context of state-of-the-art computing for the optical properties of complex dust grains. We detail the advantages of this method over existing ones and highlight what we identify as potential caveats.

 \subsection{Comparison with other methods}

As described in Sect. \ref{introduction}, OMER is not the first approximation developed to calculate the optical properties of grains with complex structures and/or compositions. Solutions have been sought to calculate the properties of grains for which, for example, no analytical solution exists and to reduce the cost of these calculations, whether approximate or exact. This section compares OMER with a few common approximations.

The \ac{emt} is a widely used approximation that simplifies the computation of grains that incorporate different materials and porous grains. In \cref{fig:aC,fig:aSil}, we compare OMER and \ac{emt} to approximate the aSil/aC and aCH/aC core-mantle grains, respectively. For silicate grains,  \ac{emt} appears to be already robust, but OMER improves the accuracy at most wavelengths. The extinction cross-section is perhaps the best example of this, with errors of $0.8\%$ averaged over all computed wavelengths for OMER, to be compared to 4.7\% for \ac{emt}. Both methods equally well approximate the scattering and backscattering cross-sections from the NIR to the centimetre range, but OMER significantly increases the accuracy in the visible range. For carbonaceous aCH/aC grains, where the core and mantle materials have highly contrasting imaginary parts of their refractive indices, \ac{emt} is inaccurate (see \cref{fig:aC}). The OMER extinction cross-section is accurate to within 1\% on average, whereas \ac{emt} is only accurate to within 15\% on average and can deviate by up to 77\% at long wavelengths. As with silicate grains, OMER performs as well as  \ac{emt} for scattering and backscattering cross-sections for $\lambda >\qty{1}{\um}$ and better in the visible with errors smaller than 15\% whereas the \ac{emt} tends to deviate by up to $>50\%$ from the reference case locally. Thus, \ac{emt} represents a barrier because of its lack of accuracy at long wavelengths for these carbonaceous grains, and there was no obvious solution to this issue. The OMER method overcomes this limitation and enables the creation of very large aCH/aC grains.

Methods have also been developed to speed up the calculation of the optical properties of aggregates. One of these is the \ac{mft} \citep[see, for instance][]{botet_sensitivity_1995}, which has been improved by \citet{tazaki_light_2018} with the \ac{mmf}. This method gives results for approximating aggregates similar to the \ac{emt} for porous aggregates, with an accuracy better than 20\% for $\lambda < \qty{20}{\um}$, but is about half as good at long wavelengths for extinction and is of the order of 30\% for scattering. More recently, \citet{botet_light_2025} refined the \ac{mmf} in the visible range and managed to achieve an accuracy approximately twice better at \qty{532}{\nm}. As shown by \citet{tazaki_light_2018}, other relatively common methods such as the distribution of hollow spheres \citep[DHS, ][]{jones_modelling_1988, min_modeling_2005} or the Rayleigh-Gans-Debye theory do not perform better when considering an extended wavelength range from the visible to the submillimetre. When OMER is applied to an aggregated grain (see \cref{fig:Agg}), the extinction cross-section is retrieved with an accuracy always better than 8\%, and for $\lambda > \qty{10}{\um}$ better than 1\%, while the error on the scattering cross-section remains below 6\%. Depending on the wavelength range, the methods mentioned previously can have accuracies equivalent to those of OMER. The particular advantage of OMER is that it maintains the accuracy of the approximation across the entire visible to millimetre range, unlike most other methods, which have 'grey areas' where accuracy is significantly lost.

\subsection{Known limits of the method}

Some grains have wavelengths for which no solution can be found using OMER (see \cref{sec:Results,sec:applications}). These problematic wavelengths appear close to $\lambda = 4a_{eff}$ and are probably related to scattering effects in porous semiconductor materials \citep[see for example][]{ysard_optical_2018, vorobiev_effects_2012}. We have not found a solution to this problem and the only advice we can give at this stage is to ignore these outliers when they appear. We emphasise that the OMER method remains an approximation method and cannot reproduce the optical properties of complex grains with perfect exactitude. In particular, we advise using the method with caution when a very precise accuracy of the spectral properties of the grains is required, for example, in the case of spectral fitting purposes. In the meantime, we also stress that such errors, which tend to happen at fixed size parameters, tend to smooth out (see \cref{app:population}) in dust populations, and overall the OMER method provides an approximation with limited errors over the whole range from the visible to submillimetre wavelengths, which makes it a useful tool for several astrophysical applications (radiative transfer simulations, for example).

The second limitation concerns the angular information contained in the optical properties (scattering, polarisation). Indeed, the principle of OMER is to simplify a ‘complicated’ grain by an equivalent sphere, which can then be used as a monomer in a large aggregate. It is therefore obvious that the transition to the sphere results in the loss of angular information. However, if we assume that the monomers are randomly oriented in an aggregate, this loss is unlikely to have any significant consequences. The angular information at the aggregate level will always be calculable. However, this remains a highly improvable aspect of the method, which will require further investigation.

This method is effective when simplifying a grain for use as a monomer, for example, but we emphasise that the OMER method can only approximate grains for which the optical properties have already been computed. The OMER method is thus useful for building large, complex aggregates; however, it cannot permit the construction of the optical properties of a grain 'from scratch', as the methods listed in the previous section can.

In any case, if the OMER method does not always result in a perfect match or one that is good enough, the method can be adapted to some extent. When optimising, the weights assigned to each of the parameters, $C_\sca, C_\ext, C_\back$, and $g$, can be adjusted to improve the accuracy of one or more of the parameters. The OMER method is built to remain flexible enough to serve the user's purposes and mitigate errors, making the OMER method rather versatile.

Finally, the method has been tested on a series of test cases that is as extensive as possible, but nonetheless remains limited. Although it provides a good approximation in the cases presented in this paper, we stress that it must be used with care by external users.

\section{Conclusions}
\label{sec:Conclusions}

In this paper, we presented a novel method for the approximation of the optical properties of dust grains using homogeneous spheres, which is promising for the computation of large aggregated dust grains with \ac{dda}. The main results of the OMER method are as follows:

\begin{enumerate}
    \item The OMER method permits us to simplify a complex monomer grain with a sophisticated structure and/or composition by means of a simple homogeneous sphere. In particular, hierarchical aggregates and mantle-covered aggregates can use OMER approximation spheres as monomers.\\
    \item The method provides an accurate representation of \ac{cm} grains, the relative error of OMER is $<15\%$ for the scattering and extinction cross-sections, and drops below $2\%$ for extinction at $\lambda>\qty{1}{\um}$. \\
    \item For aggregated grains, aggregates made of OMER monomers reproduce the optical properties of the exact aggregate with great accuracy ($<7\%$) from the visible to the sub-millimetre range.\\
    \item In practical cases, the OMER grains remain fairly consistent, with errors lower than $20\%$, and allow considerable reductions in computing time up to several orders of magnitude. Although caution must be paid to results fluctuations in the case of spectral fitting purposes, the method provides good accuracy regardless of the wavelength range of interest, making it a particularly valuable tool for radiative transfer simulation of particle populations.
\end{enumerate}

\begin{acknowledgements}
This work was made possible thanks to the support from the European Research Council (ERC) under the European Union’s Horizon 2020 research and innovation programme (Grant agreement No. 101098309 - PEBBLES).
Colour sequences in plots of this project are designed to be accessible thanks to \citet{petroff_accessible_2021}. A.M. acknowledges support by the program Unidad de Excelencia María de
Maeztu, awarded to the Institut de Ciències de l’Espai (CEX2020-001058-M).
\end{acknowledgements}

\bibliographystyle{aa} 
\bibliography{bibliography} 

@article{hertz_ueber_1882,
  title = {{Ueber die Ber\"uhrung fester elastischer K\"orper.}},
  author = {Hertz, Heinrich},
  year = 1882,
  journal = {Journal f\"ur die reine und angewandte Mathematik},
  volume = {92},
  pages = {156--171},
  langid = {german}
}

@article{dominik_physics_1997,
  title = {The {{Physics}} of {{Dust Coagulation}} and the {{Structure}} of {{Dust Aggregates}} in {{Space}}},
  author = {Dominik, C. and Tielens, A. G. G. M.},
  year = 1997,
  month = may,
  journal = {The Astrophysical Journal},
  volume = {480},
  number = {2},
  pages = {647},
  issn = {0004-637X},
  doi = {10.1086/303996}
}

@book{bohren_absorption_1983,
  title = {Absorption and Scattering of Light by Small Particles},
  author = {Bohren, Craig F. and Huffman, Donald R.},
  year = 1983,
  publisher = {J. Wiley \& sons},
  address = {New York Chichester Brisbane [Australia] [etc.]},
  isbn = {978-0-471-05772-7},
  langid = {english},
  lccn = {535.3}
}

@article{botet_light_2025,
  title = {Light Scattering by Large Particles -- a Refinement to the {{Mean-Field T-matrix}} Method --},
  author = {Botet, Robert},
  year = 2025,
  journal = {Journal of Quantitative Spectroscopy and Radiative Transfer},
  volume = {347},
  pages = {109589},
  issn = {0022-4073},
  doi = {10.1016/j.jqsrt.2025.109589}
}

@article{botet_sensitivity_1995,
  title = {Sensitivity of Some Optical Properties of Fractals to the Cut-off Functions},
  author = {Botet, R. and Rannou, P. and Cabane, M.},
  year = 1995,
  month = jan,
  journal = {Journal of Physics A Mathematical General},
  volume = {28},
  number = {2},
  pages = {297--316},
  doi = {10.1088/0305-4470/28/2/008}
}

@article{bruggeman_berechnung_1935,
  title = {Berechnung Verschiedener Physikalischer {{Konstanten}} von Heterogenen {{Substanzen}}. {{I}}. {{Dielektrizit\"atskonstanten}} Und {{Leitf\"ahigkeiten}} Der {{Mischk\"orper}} Aus Isotropen {{Substanzen}}},
  author = {Bruggeman, D. A. G.},
  year = 1935,
  journal = {Annalen der Physik},
  volume = {416},
  number = {7},
  pages = {636--664},
  doi = {10.1002/andp.19354160705}
}

@article{carpine_small_2025,
  title = {From Small Dust to Micron-Sized Aggregates: {{The}} Influence of Structure and Composition on the Dust Optical Properties},
  author = {Carpine, M.-A. and Ysard, N. and Maury, A. and Jones, A.},
  year = 2025,
  journal = {\aap},
  volume = {698},
  pages = {A200},
  doi = {10.1051/0004-6361/202554575}
}

@article{doner_impact_2017,
  title = {Impact of Morphology on the Radiative Properties of Fractal Soot Aggregates},
  author = {Doner, Nimeti and Liu, Fengshan},
  year = 2017,
  journal = {\jqsrt},
  volume = {187},
  pages = {10--19},
  issn = {0022-4073},
  doi = {10.1016/j.jqsrt.2016.09.005}
}

@article{draine_discrete-dipole_1988,
  title = {The Discrete-Dipole Approximation and Its Application to Interstellar Graphite Grains},
  author = {Draine, B. T.},
  year = 1988,
  month = oct,
  journal = {\apj},
  volume = {333},
  pages = {848},
  doi = {10.1086/166795}
}

@article{draine_spheroidal_2025,
  title = {The Spheroidal Analog Method for Modeling Irregular Porous Aggregates},
  author = {Draine, B. T.},
  year = 2025,
  month = may,
  journal = {\apj},
  volume = {985},
  number = {1},
  pages = {10},
  issn = {1538-4357},
  doi = {10.3847/1538-4357/adc57d}
}

@article{garnett_xii_1904,
  title = {{{XII}}. {{Colours}} in Metal Glasses and in Metallic Films},
  author = {Garnett, J. C. Maxwell},
  year = 1904,
  month = jan,
  journal = {Philosophical Transactions of the Royal Society of London, Series A},
  volume = {203},
  number = {359-371},
  pages = {385--420},
  issn = {0264-3952},
  doi = {10.1098/rsta.1904.0024}
}

@article{jaeger_steps_1998,
  title = {Steps toward Interstellar Silicate Mineralogy. {{IV}}. {{The}} Crystalline Revolution},
  author = {Jaeger, C. and Molster, F. J. and Dorschner, J. and Henning, {\relax Th}. and Mutschke, H. and Waters, L. B. F. M.},
  year = 1998,
  month = nov,
  journal = {\aap},
  volume = {339},
  pages = {904--916}
}

@article{jones_modelling_1988,
  title = {Modelling Interstellar Extinction -- {{I}}. {{Porous}} Grains},
  author = {Jones, A. P.},
  year = 1988,
  month = sep,
  journal = {\mnras},
  volume = {234},
  number = {2},
  pages = {209--218},
  issn = {0035-8711},
  doi = {10.1093/mnras/234.2.209}
}

@article{jones_spheroidal_2026,
  title = {Spheroidal Core-Mantle Particle Absorption, Scattering, and Polarisation in the Long-Wavelength Limit},
  author = {Jones, A.P. and Ysard, N.},
  year = 2026,
  month = jan,
  journal = {\aap},
  volume = {706},
  pages = {A250},
  issn = {0004-6361, 1432-0746},
  doi = {10.1051/0004-6361/202557518},
  urldate = {2026-01-16},
  copyright = {https://www.edpsciences.org/en/authors/copyright-and-licensing}
}

@incollection{kimura_light_2016,
  title = {Light Scattering and Thermal Emission by Primitive Dust Particles in Planetary Systems},
  booktitle = {Light Scattering Reviews, Volume 11},
  author = {Kimura, Hiroshi and Kolokolova, Ludmilla and Li, Aigen and Lebreton, J{\'e}r{\'e}my},
  year = 2016,
  pages = {363--418},
  publisher = {Springer Berlin Heidelberg},
  isbn = {978-3-662-49538-4}
}

@article{kohler_aggregate_2011,
  title = {Aggregate Dust Connections and Emissivity Enhancements},
  author = {K{\"o}hler, M. and Guillet, V. and Jones, A.},
  year = 2011,
  month = apr,
  journal = {\aap},
  volume = {528},
  pages = {A96},
  doi = {10.1051/0004-6361/201016379}
}

@article{kohler_dust_2015,
  title = {Dust Evolution in the Transition towards the Denser {{ISM}}: Impact on Dust Temperature, Opacity, and Spectral Index},
  author = {K{\"o}hler, M. and Ysard, N. and Jones, A. P.},
  year = 2015,
  month = jul,
  journal = {\aap},
  volume = {579},
  pages = {A15},
  doi = {10.1051/0004-6361/201525646}
}

@article{kozasa_optical_1992,
  title = {Optical Properties of Dust Aggregates. {{I}} - {{Wavelength}} Dependence},
  author = {Kozasa, Takashi and Blum, Juergen and Mukai, Tadashi},
  year = 1992,
  month = sep,
  journal = {\aap},
  volume = {263},
  number = {1-2},
  pages = {423--432}
}

@article{lorek_local_2018,
  title = {Local Growth of Dust- and Ice-Mixed Aggregates as Cometary Building Blocks in the Solar Nebula},
  author = {Lorek, S. and Lacerda, P. and Blum, J.},
  year = 2018,
  journal = {\aap},
  volume = {611},
  pages = {A18},
  doi = {10.1051/0004-6361/201630175}
}

@article{min_modeling_2005,
  title = {Modeling Optical Properties of Cosmic Dust Grains Using a Distribution of Hollow Spheres},
  author = {Min, M. and Hovenier, J. W. and {de Koter}, A.},
  year = 2005,
  month = mar,
  journal = {\aap},
  volume = {432},
  number = {3},
  pages = {909--920},
  doi = {10.1051/0004-6361:20041920}
}

@article{min_multiwavelength_2016,
  title = {Multiwavelength Optical Properties of Compact Dust Aggregates in Protoplanetary Disks},
  author = {Min, M. and Rab, {\relax Ch}. and Woitke, P. and Dominik, C. and M{\'e}nard, F.},
  year = 2016,
  journal = {\aap},
  volume = {585},
  pages = {A13},
  doi = {10.1051/0004-6361/201526048}
}

@article{min_shape_2003,
  title = {Shape Effects in Scattering and Absorption by Randomly Oriented Particles Small Compared to the Wavelength},
  author = {Min, M. and Hovenier, W., J. and {de Koter}, A.},
  year = 2003,
  journal = {\aap},
  volume = {404},
  number = {1},
  pages = {35--46},
  doi = {10.1051/0004-6361:20030456}
}

@article{mishchenko_applicability_2016,
  title = {Applicability of the Effective-Medium Approximation to Heterogeneous Aerosol Particles},
  author = {Mishchenko, M. I. and Dlugach, J. M. and Liu, L.},
  year = 2016,
  journal = {\jqsrt},
  volume = {178},
  pages = {284--294},
  doi = {10.1016/j.jqsrt.2015.12.028}
}

@article{mishchenko_t-matrix_1996,
  title = {T-Matrix Computations of Light Scattering by Nonspherical Particles: {{A}} Review},
  author = {Mishchenko, Michael I. and Travis, Larry D. and Mackowski, Daniel W.},
  year = 1996,
  journal = {\jqsrt},
  volume = {55},
  number = {5},
  pages = {535--575},
  issn = {0022-4073},
  doi = {10.1016/0022-4073(96)00002-7}
}

@article{okamoto_light_1998,
  title = {Light Scattering by Irregular Interplanetary Dust Particles},
  author = {Okamoto, Hajime and Xu, Yu-lin},
  year = 1998,
  month = jun,
  journal = {Earth, Planets and Space},
  volume = {50},
  number = {6-7},
  pages = {577--585},
  doi = {10.1186/BF03352151}
}

@article{ormel_dust_2009,
  title = {Dust Coagulation and Fragmentation in Molecular Clouds. {{I}}. {{How}} Collisions between Dust Aggregates Alter the Dust Size Distribution},
  author = {Ormel, C. W. and Paszun, D. and Dominik, C. and Tielens, A. G. G. M.},
  year = 2009,
  month = aug,
  journal = {\aap},
  volume = {502},
  number = {3},
  pages = {845--869},
  doi = {10.1051/0004-6361/200811158}
}

@article{petroff_accessible_2021,
  title = {Accessible {{Color Sequences}} for {{Data Visualization}}},
  author = {Petroff, Matthew A.},
  year = 2021,
  journal = {arXiv e-prints},
  pages = {arXiv:2107.02270},
  doi = {10.48550/ARXIV.2107.02270},
  urldate = {2024-12-12},
  copyright = {arXiv.org perpetual, non-exclusive license}
}

@misc{prahl_miepython_2025,
  title = {Miepython: {{Pure}} Python Calculation of {{Mie}} Scattering},
  author = {Prahl, Scott},
  year = 2025,
  month = may,
  howpublished = {Zenodo}
}

@article{purcell_scattering_1973,
  title = {Scattering and Absorption of Light by Nonspherical Dielectric Grains},
  author = {Purcell, Edward M. and Pennypacker, Carlton R.},
  year = 1973,
  month = dec,
  journal = {\apj},
  volume = {186},
  pages = {705--714},
  doi = {10.1086/152538}
}

@article{sumlin_retrieving_2018,
  title = {Retrieving the Aerosol Complex Refractive Index Using {{PyMieScatt}}: {{A Mie}} Computational Package with Visualization Capabilities},
  author = {Sumlin, Benjamin J. and Heinson, William R. and Chakrabarty, Rajan K.},
  year = 2018,
  journal = {\jqsrt},
  volume = {205},
  pages = {127--134},
  issn = {0022-4073},
  doi = {10.1016/j.jqsrt.2017.10.012}
}

@article{tazaki_light_2018,
  title = {Light {{Scattering}} by {{Fractal Dust Aggregates}}. {{II}}. {{Opacity}} and {{Asymmetry Parameter}}},
  author = {Tazaki, Ryo and Tanaka, Hidekazu},
  year = 2018,
  month = jun,
  journal = {\apj},
  volume = {860},
  number = {1},
  pages = {79},
  publisher = {American Astronomical Society},
  issn = {1538-4357},
  doi = {10.3847/1538-4357/aac32d}
}

@article{vorobiev_effects_2012,
  title = {The Effects of Porosity on Optical Properties of Semiconductor Chalcogenide Films Obtained by the Chemical Bath Deposition},
  author = {Vorobiev, Yuri V. and Horley, Paul P. and {Hern{\'a}ndez-Borja}, Jorge and {Esparza-Ponce}, Hilda E. and {Ram{\'\i}rez-Bon}, Rafael and Vorobiev, Pavel and P{\'e}rez, Claudia and {Gonz{\'a}lez-Hern{\'a}ndez}, Jes{\'u}s},
  year = 2012,
  month = dec,
  journal = {Nanoscale Research Letters},
  volume = {7},
  number = {1},
  pages = {483},
  doi = {10.1186/1556-276X-7-483}
}

@article{ysard_optical_2018,
  title = {The Optical Properties of Dust: The Effects of Composition, Size, and Structure},
  author = {Ysard, N. and {Jones, A. P.} and {Demyk, K.} and {Bout\'eraon, T.} and {Koehler, M.}},
  year = 2018,
  journal = {\aap},
  volume = {617},
  pages = {A124},
  doi = {10.1051/0004-6361/201833386}
}

@article{ysard_themis_2024,
  title = {{{THEMIS}} 2.0: {{A}} Self-Consistent Model for Dust Extinction, Emission, and Polarisation},
  author = {Ysard, N. and {Jones, A. P.} and {Guillet, V.} and {Demyk, K.} and {Decleir, M.} and {Verstraete, L.} and {Choubani, I.} and {Miville-Desch\^enes, M.-A.} and {Fanciullo, L.}},
  year = 2024,
  journal = {\aap},
  volume = {684},
  pages = {A34},
  doi = {10.1051/0004-6361/202348391}
}

@article{yurkin_discrete-dipole-approximation_2011,
  title = {The Discrete-Dipole-Approximation Code {{ADDA}}: {{Capabilities}} and Known Limitations},
  author = {Yurkin, Maxim A. and Hoekstra, Alfons G.},
  year = 2011,
  journal = {\jqsrt},
  volume = {112},
  number = {13},
  pages = {2234--2247},
  issn = {0022-4073},
  doi = {10.1016/j.jqsrt.2011.01.031}
}
\FloatBarrier
~
\FloatBarrier

\begin{appendix}
\section{Effects of size}
\label{app:size}

\begin{figure}[H]
\centering
\begin{subfigure}{\linewidth}
  \centering
  \subfigimg[width=\linewidth,clip]{pos=ur, hsep = \linewidth-2.5cm, vsep =1cm}{\small (a)}{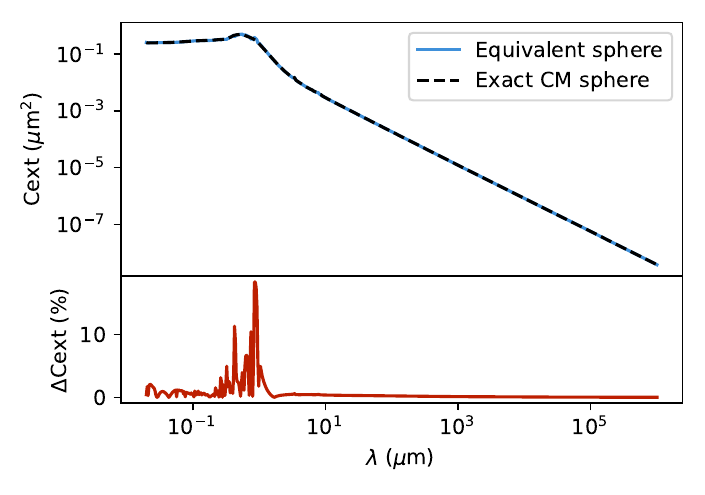}
  \phantomsubcaption
  \label{fig:aC200(a)}
\end{subfigure}%
\newline
\begin{subfigure}{\linewidth}
  \centering
  \subfigimg[width=\linewidth,clip]{pos=ur, hsep = \linewidth-2.5cm, vsep =1cm}{\small (b)}{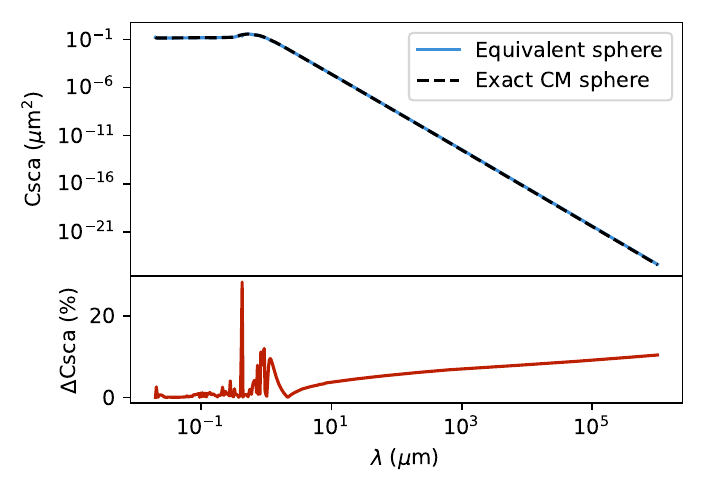}
  \phantomsubcaption
  \label{fig:aC200(b)}
\end{subfigure}
\caption{Optical properties of a \qty{200}{\nm} grain with an aCH core and a \qty{20}{\nm} aC mantle (black dashed line) and the equivalent representation of the same grain with OMER (blue): extinction (a) and scattering (b) cross-sections. Bottom panels show the absolute value of the relative difference between the exact \texttt{miepython} computation and the equivalent representations.}
\label{fig:aC200}
\end{figure}
\FloatBarrier 

In this section we study the effect of grain size on the wavelength at which the errors on the cross-section peaks.
First, we derive the same results as the one for the $a_0 = \qty{100}{\nm}$ aC/aCH grain described in Sect. \ref{subsec:CM} for a larger size of $a_0 = \qty{200}{nm}$. The results are shown in \cref{fig:aC200}. The order of magnitude of the errors remains the same as in the case of the smaller grain, showing not dependency of the OMER performance on the grain size. Interestingly, we note that the error peak that was seen in \cref{fig:aC} around $4 a_0 = \qty{400}{\nm}$ is present as well but here around $4 a_0 = \qty{800}{\nm}$, showing that the error peak is related to scattering effects and can be predictable.

\begin{figure}[H]
\centering
\begin{subfigure}{\linewidth}
  \centering
  \subfigimg[width=\linewidth,clip]{pos=ur, hsep = \linewidth-2.5cm, vsep =1cm}{\small (a)}{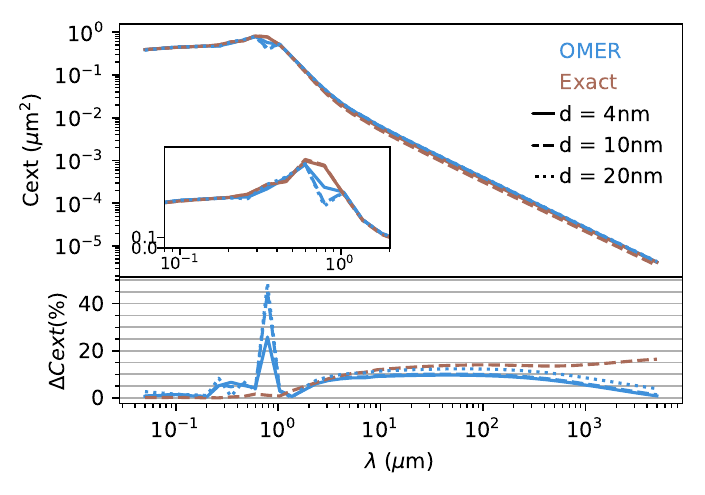}
  \phantomsubcaption
  \label{fig:Double200(a)}
\end{subfigure}%
\newline
\begin{subfigure}{\linewidth}
  \centering
  \subfigimg[width=\linewidth,clip]{pos=ur, hsep = \linewidth-2.5cm, vsep =1cm}{\small (b)}{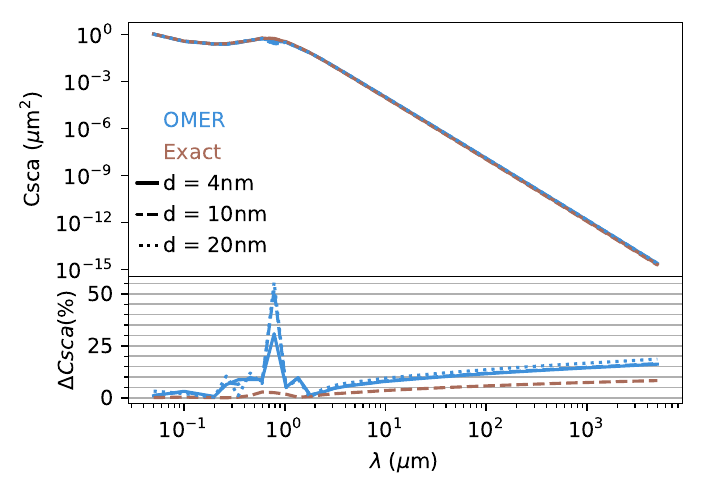}
  \phantomsubcaption
  \label{fig:Double200(b)}
\end{subfigure}
\caption{Optical properties of an aggregate composed of two 200nm spherical monomers with aCH cores and 20 nm-thick aC mantles (brown) and of the equivalent representation of the same grain with OMER spheres for monomers (blue): extinction (a) and scattering (b) cross-sections. Computations are made with several dipole sizes $d=\qty{4}{nm}$ (solid lines), $d=\qty{10}{nm}$ (dashed), and  $d=\qty{20}{nm}$ (dotted). Bottom panels show the absolute value of the relative difference between the exact \ac{dda} computation (core-mantle with $d=\qty{4}{nm}$) and the other representations. In panel (a), a zoom-in of the optical part of the curve is plotted with linear y-scale.}
\label{fig:Double200}
\end{figure}

Then, we plot in \cref{fig:Double200} the same results as in \cref{fig:2sph}, but with a double sphere composed of \qty{200}{\nm} aC/aCH monomers. We do notice a similar error peak in the resulting optical properties. Similarly as with the core-mantle single sphere, the error peak follows the grain size: if the error was maximal at $\lambda=\qty{400}{\nm}$ for the \qty{100}{\nm} spheres, here the error peaks at \qty{800}{\nm}. We do notice, once again, that those error peaks are intrinsically linked to the size of the grains, even when the OMER grains are incorporated in larger structures.

\FloatBarrier

\FloatBarrier

\section{Effects of elongation}
\label{app:spheroid}
\FloatBarrier
\begin{figure}[h]
\centering
\begin{subfigure}{\linewidth}
  \centering
  \subfigimg[width=\linewidth]{pos=ur, hsep = \linewidth-2.5cm, vsep =1cm}{\small (a)}{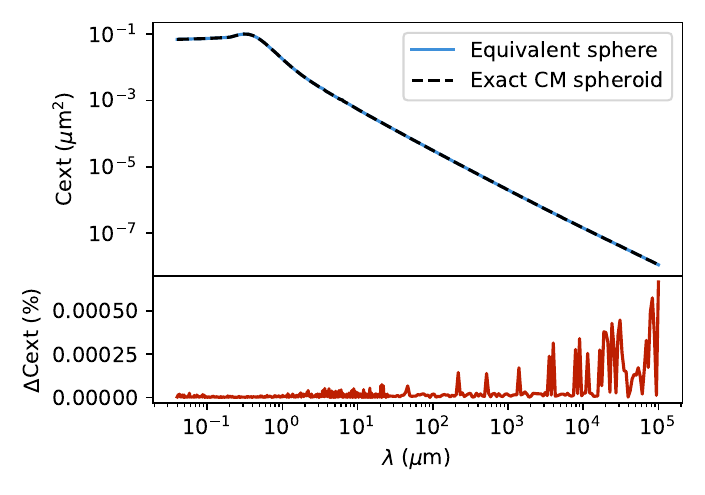}
  \phantomsubcaption
  \label{fig:spheroid(a)}
\end{subfigure}%
\newline
\begin{subfigure}{\linewidth}
  \centering
  \subfigimg[width=\linewidth]{pos=ur, hsep = \linewidth-2.5cm, vsep =1cm}{\small (b)}{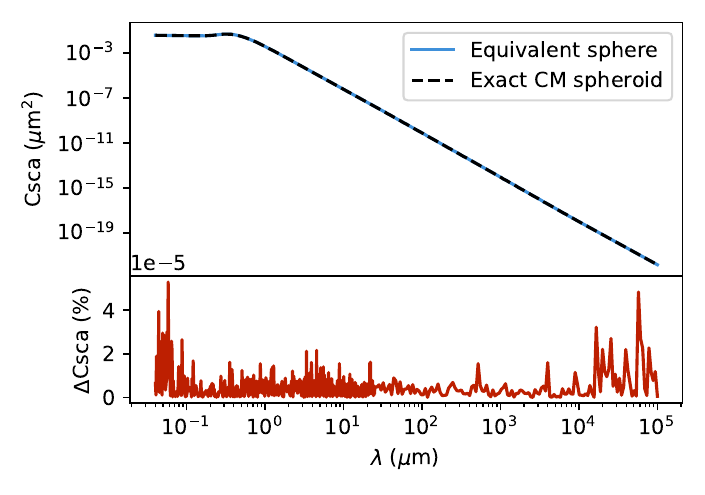}
  \phantomsubcaption
  \label{fig:spheroid(b)}
\end{subfigure}%
\caption{Optical properties of a \qty{100}{\nm} spheroidal (prolate, aspect ratio 0.5) grain with an aCH core and \qty{20}{\nm} aC mantle (black dashed line), and the equivalent representation of the same grain with OMER (blue): extinction (a) and scattering (b) cross-sections. Bottom panels show the absolute value of the relative difference between the exact \ac{dda} computation and the equivalent representations.}
\label{fig:spheroid}
\end{figure}

In this section we derive the same results as the one for the \qty{100}{\nm} aC/aCH grain described in Sect. \ref{subsec:CM} but departing from the sphere by using a prolate spheroid of aspect ratio 0.5. Note that, in this case, the optimisation is run only on $C_\sca$ and $C_\ext$ due to lack of data: this grain is taken from the THEMIS 2.0 grain model, readily computed, for which the public data include extinction and scattering parameters. Besides, we take this as an opportunity to try the method on a reduced set of optimisation parameters. We plot the results for the extinction and scattering cross-sections in \cref{fig:spheroid}. The order of magnitude of the errors remains the same as in the case of the spherical grain, showing that the OMER accuracy is not affected by the particular elongation of the grain and is able to reproduce the increase in the extinction cross-section expected for aspect ratios not equal to 1; in the mean time, the method proves its versatility in using a reduced set of parameters for optimisation.

\section{Crystalline material}
\label{app:crystal}

\FloatBarrier 

\begin{figure}[H]
\centering
\begin{subfigure}{\linewidth}
  \centering
  \subfigimg[width=\linewidth,clip]{pos=ur, hsep = \linewidth-2.5cm, vsep =1cm}{\small (a)}{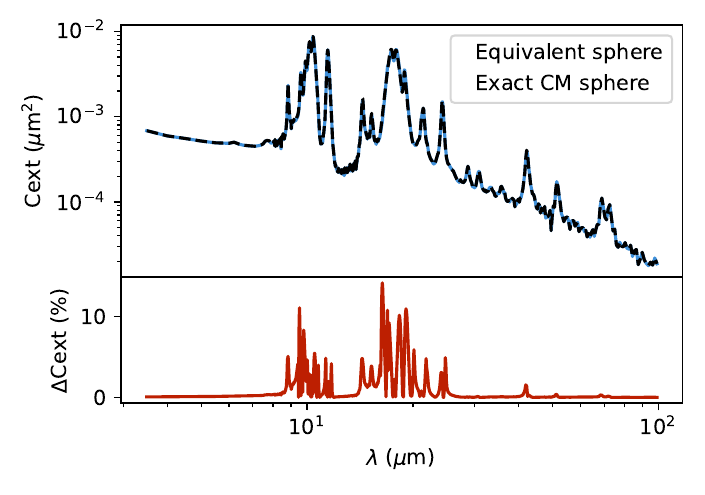}
  \phantomsubcaption
  \label{fig:cryst(a)}
\end{subfigure}%
\newline
\begin{subfigure}{\linewidth}
  \centering
  \subfigimg[width=\linewidth,clip]{pos=ur, hsep = \linewidth-2.5cm, vsep =1cm}{\small (b)}{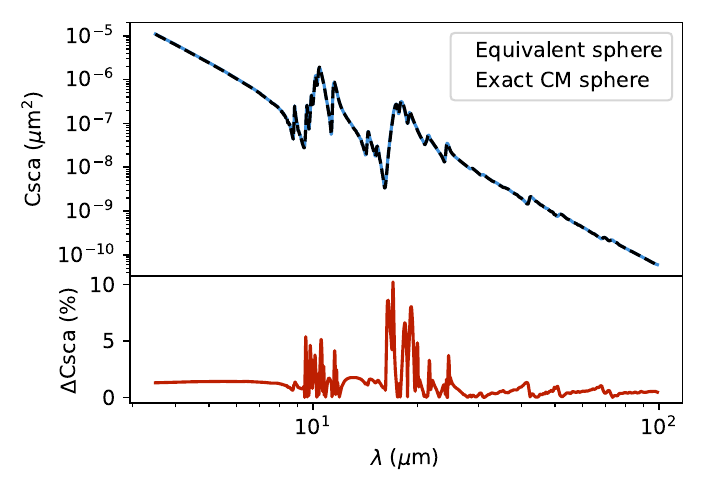}
  \phantomsubcaption
  \label{fig:cryst(b)}
\end{subfigure}
\caption{Optical properties of a \qty{100}{\nm} grain with a crystalline Silicate core and a \qty{5}{\nm} aC mantle (black dashed line) and the equivalent representation of the same grain with OMER (blue): extinction (a) and scattering (b) cross-sections. Bottom panels show the absolute value of the relative difference between the exact \texttt{miepython} computation and the equivalent representations.}
\label{fig:crystal}
\end{figure}

In this section we derive the same results as the one for the \qty{100}{\nm} aSil/aCH grain described in Sect. \ref{subsec:CM}, but with crystalline silicate material from instead of the aSil core: we use the Enstatite material from \citet{jaeger_steps_1998}. The results are shown in \cref{fig:crystal}. The order of magnitude of the errors remains the same as in the case of the purely amorphous grain, showing that the OMER method is capable of keeping intact the fine cristalline features in a grain.

\FloatBarrier

\section{Grain populations}
\label{app:population}

\FloatBarrier 

\begin{figure}
    \centering
    \includegraphics[width=\linewidth]{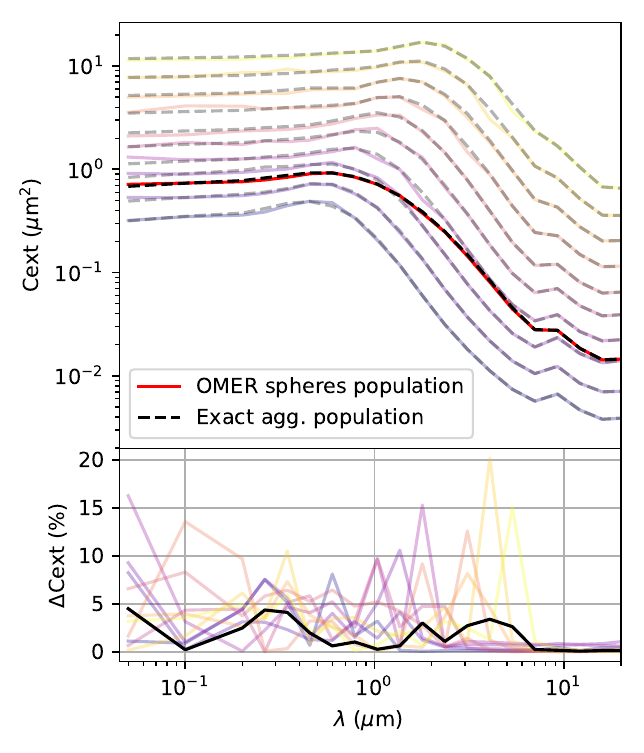}
    \caption{Extinction cross-section of a selection of 10 aggregated grain (light-shaded curves) made of aCH/aC and aSil/aC core-mantle \qty{100}{\nm} spherical monomers (dashed black) and the equivalent representation of the same grain with OMER (color). Total sizes of aggregates are logarithmically sampled from \qty{200}{\nm} (dark purple) to \qty{1}{\um} (light yellow). In bright-colored (black and red), we overplot the extinction cross-section of a power-law ($q=-3.5$) distributed population of those grains, both exact (dashed black), and OMER (red). Bottom panel represents the absolute value of the relative difference between the exact \ac{dda} computation and the OMER equivalent representation, for the inidividual grains (light shaded, color) and the dust population (black).}
    \label{fig:Population}
\end{figure}

In this section we propose a simple case to test the behaviour of the OMER method in a dust population including several grain sizes. We reproduced the case of approximation of aggregates with OMER (Sect. \ref{subsec:agg}), over a size selection in the range $[\qty{200}{nm};\qty{1}{\um}]$. In particular, we produced the OMER equivalent sphere of 10 aggregated grains of different size, made out of $\qty{100}{nm}$ spherical monomers. For each case (both exact and OMER-approximated), we also combined the grains into a single population, following a power-law size distribution ($q=-3.5$), by averaging the extinction cross-section over all sizes. The results are presented in \cref{fig:Population}. We do observe the averaging of errors in the population: if in individual grains, the errors on $C_\ext$ peak at values of $10\%$ or, for some grains, up to $20\%$, the population displays errors which always remain $<5\%$ (outside of the plotted range, for $\lambda>\qty{10}{\um}$ the errors on individual grains are all $<2\%$ and on the population $<0.4\%)$). We note that those errors are smoothed in the population, and particularly, that besides being smaller in absolute value, they are less sharp. Overall, we also emphasise that the OMER method provides low errors on the whole range of wavelengths tested, which makes it stand out from the existing method listed in Sect. \ref{introduction}.

\FloatBarrier

\end{appendix}

\end{document}